\documentclass[aps,prd,superscriptaddress,floatfix,nofootinbib,showpacs]{revtex4}
\usepackage{subfigure}
\usepackage{color}
\usepackage[dvips]{graphicx}

\newcommand{\lsim}{\mathrel{\mathop{\kern 0pt \rlap
      {\raise.2ex\hbox{$<$}}}\lower.9ex\hbox{\kern-.190em $ \sim$}}}

\newcommand{\gsim}{\mathrel{\mathop{\kern 0pt
      \rlap{\raise.2ex\hbox{$>$}}}\lower.9ex\hbox{\kern-.190em $\sim
      $}}}

\newcommand{\beq}    {\begin{equation}}
\newcommand{\eeq}    {\end{equation}}

\newcommand{\be}{\begin{equation}}
\newcommand{\ee}{\end{equation}}
\newcommand{\beqarr}{\begin{eqnarray}}
\newcommand{\eeqarr}{\end{eqnarray}}

\newcommand{\msq}{$m_{\tilde q}$}
\newcommand{\mll}{$m_{\ell\ell}$}
\newcommand{\mjll}{$m_{j\ell\ell}$}
\newcommand{\mjllo}{$m_{j\ell(lo)}$}
\newcommand{\mjlhi}{$m_{j\ell(hi)}$}

\begin{document}

\title{Search at the CERN LHC for a light neutralino of cosmological
  interest }

\thanks{Preprint number: DFTT 21/2011}


\author{S. Choi}
\affiliation{Korea University, Seoul, Korea, 136-701}
\author{S. Scopel}
\affiliation{Department of Physics, Sogang University\\
Seoul, Korea, 121-742}
\author{N. Fornengo}
\affiliation{Dipartimento di Fisica Teorica, Universit\`a di Torino \\
Istituto Nazionale di Fisica Nucleare, Sezione di Torino \\
via P. Giuria 1, I--10125 Torino, Italy}
\author{A. Bottino}
\affiliation{Dipartimento di Fisica Teorica, Universit\`a di Torino \\
Istituto Nazionale di Fisica Nucleare, Sezione di Torino \\
via P. Giuria 1, I--10125 Torino, Italy}
\date{\today}

\begin{abstract}
We address the problem of a search at the LHC for a neutralino whose
mass is around 10 GeV, i.e. in the range of interest for present data
of direct search for dark matter particles in the galactic halo.  This
light neutralino is here implemented in an effective Minimal
Supersymmetric extension of the Standard Model at the electroweak
scale without requirement of a gaugino-mass unification at a grand
unification scale.  Within this model we select a representative
benchmark and determine its prospects of reconstructing the main
features of the model at different stages of the LHC runs.
\end{abstract}


\maketitle

\section{Introduction}
\label{sec:intro}

Most investigations on the search for neutralinos at the CERN Large
Hadron Collider (LHC) concern neutralinos within
Supergravity--inspired (SUGRA) models. This implies, in particular,
neutralinos of a mass larger than about 50 GeV, since this is the
lower bound on the neutralino mass which directly follows from the LEP
lower bound on the chargino mass combined with the assumption,
inherent in SUGRA models, that the gaugino masses are unified at a
grand unification (GUT) scale.

Relaxation of this hypothesis of gaugino-mass unification allows the
neutralino mass $m_{\chi}$ to be smaller than 50 GeV. A supersymmetric
model which incorporates this possibility and is still very
manageable, since expressible in terms of a limited number of
independent parameter, is the one considered in Ref.~\cite{lowneu}. It
consists of an effective Minimal Supersymmetric extension of the
Standard Model (MSSM) at the electroweak (EW) scale, whose main
properties are summarized in Sect. \ref{sec:model}.

In Ref.~\cite{lowneu} it was stressed that, in case of R-parity
conservation, a light neutralino ({\it i.e.} a neutralino with
$m_{\chi} \lsim$ 50 GeV), when happens to be the Lightest
Supersymmetric Particle (LSP), constitutes an extremely interesting
candidate for the dark matter (DM) in the Universe, with direct
detection rates accessible to experiments of present generation. 

In Ref.~\cite{lowneu} it was also derived a lower bound for $m_{\chi}$
from the cosmological upper limit on the cold dark matter density; the
value of this lower bound, updated on the basis of the experimental
data available in Autumn 2010, was established in
Ref. \cite{noi_discussing} to be $m_{\chi} \sim$ 7.5 GeV.  The
implementation of the very recent upper bound on the branching ratio
for the process $BR(B_s^{0} \rightarrow \mu^{-} + \mu^{+})$
\cite{new_Bs} moves now the lower bound on the neutralino mass to the
value $m_{\chi} \sim$ 9 GeV.  The theoretical framework which allows
neutralinos with a mass in the range 9 GeV $\lsim m_{\chi} \lsim$ 50
GeV is briefly summarized in Sect. II and will be simply denominated
as Light Neutralino Model (LNM) in the present paper.

In Ref.~\cite{bdfs2004} the LNM was proved to fit the annual
modulation effect measured by the DAMA collaboration
\cite{dama2003}. Our model is also compatible with all experimental
searches for indirect evidence of SUSY and with precision data that
set constraints on possible effects due to supersymmetry, as discussed
in detail in Ref.\cite{noi_discussing} (for the compatibility of very
light neutralino masses with various laboratory bounds see also
Ref. \cite{Dreiner}).  Moreover, the possible impact of some early
analyses by the CMS and ATLAS Collaborations at the LHC on the LNM was
investigated in Ref. \cite{noi_impact}.

At the same time much interest has recently been raised by a new
measurement of an annual modulation effect by the CoGeNT
Collaboration~\cite{cogent} and by some hints of possible signals of
dark matter (DM) particles in other experiments of direct detection
(CDMS \cite{cdms}, CRESST \cite{cresst}). 

What is intriguing in all the experimental results listed above is
that, if actually due to a Weakly Interacting Massive Particle (WIMP)
with a coherent interaction with the atomic nuclei of the detector
material, they would all be explained by a WIMP physical region with a
light mass (around 10 GeV) and a nucleon elastic cross--section in
agreement with the intervals for the same parameters established by
the DAMA Collaboration from a measurement of the annual modulation
that has now reached a high statistical significance by a running over
13 yearly cycles with the DAMA/NaI and the DAMA/LIBRA experiments
\cite{dama2010}.  Specifically, compatibility of the DAMA result with
CDMS has been discussed in Ref.~\cite{noi_cdms}, and between CoGeNT
and DAMA in Ref.~\cite{noi_belli_et_al_2011}.

These results have prompted a large number of phenomenological papers
focused on WIMPs with a light mass~\cite{fossa_comune}. Turning to a
specific candidate, it has now become common to consider neutralinos
with a mass of order 10 GeV.

Due to the relevant role that light neutralinos can have in cosmology
and astrophysics it becomes so of the upmost interest to investigate
the possibility of searching for these particles at the LHC.  A
preliminary analysis in this direction was performed in
Ref.~\cite{lhc1}; event rates were determined in specific scenarios
and benchmarks dictated by the relevant cosmological properties of the
LNM. There it was shown that the perspectives of a fruitful
investigation of the supersymmetric parameter space relevant for light
neutralinos at the LHC are potentially good, though no specific
analyses of the signal/background ratios and of kinematical
distributions were performed.

In the present paper we wish to extend the investigation of
Ref.~\cite{lhc1} by making use of a numerical simulation to estimate
in a realistic way the detectability of the LNM at the LHC over the SM
background and to show what information about the masses of SUSY
particles can be extracted from the data.

The paper is organized as follows. In Section \ref{sec:model} we
summarize the main features of the LNM, concentrating on the particle
mass spectra relevant for our analysis. In Section \ref{sec:signals}
we discuss the general properties in the LNM of the decay chains that
are used to reconstruct the SUSY masses.  In Section
\ref{sec:selecting_benchmarks} we explore the LNM parameter space to
select a suitable benchmark for our scenario, that is then used in
Section \ref{seq:early} to assess its chances for an early discovery
within the 7 TeV run at the LHC, and in Section \ref{sec:spectroscopy}
to discuss how the SUSY masses can be reconstructed from
invariant--mass spectra measured with a larger collected luminosity at
14 TeV. We then give our conclusions in Section \ref{sec:conclusions}.

\section{A model for light neutralinos (LNM)}
\label{sec:model}

The supersymmetric scheme we employ in the present paper is an
effective MSSM scheme (effMSSM) at the electroweak scale, with the
following independent parameters: $M_1, M_2, M_3, \mu, \tan\beta, m_A,
m_{\tilde q}, m_{\tilde{t}}, m_{\tilde l}$ and $A$.  Notations are as
follows: $M_1$, $M_2$ and $M_3$ are the U(1), SU(2) and SU(3) gaugino
masses (these parameters are taken here to be positive), $\mu$ is the
Higgs mixing mass parameter, $\tan\beta$ the ratio of the two Higgs
v.e.v.'s, $m_A$ the mass of the CP-odd neutral Higgs boson, $m_{\tilde
  q}$ is a squark soft--mass common to the squarks of the first two
families, $m_{\tilde t}$ is the squark soft--mass of the third family,
$m_{\tilde l}$ is a slepton soft--mass common to all sleptons, and $A$
is a common dimensionless trilinear parameter for the third family,
$A_{\tilde b} = A_{\tilde t} \equiv A m_{\tilde t}$ and $A_{\tilde
  \tau} \equiv A m_{\tilde l}$ (the trilinear parameters for the other
families being set equal to zero).  In our model, no gaugino mass
unification at a Grand Unified (GUT) scale is assumed: this implies
that $M_1$ and $M_2$ are independent parameters at the EW scale. The
model introduced here is the one discussed in Ref. \cite{lowneu}, with
the minimal extension that the degeneracy between the soft squark mass
of the first two families and that of the third family is removed.  In
particular, the splitting between $m_{\tilde q}$ and $m_{\tilde t}$
reduces some tuning introduced in the parameter space by the
constraint on the $b \rightarrow s + \gamma$ when $m_{\tilde
  t}$=$m_{\tilde q}\gsim$ 700 GeV~\cite{noi_impact}.

The following experimental constraints are imposed: accelerators data
on supersymmetric and Higgs boson searches (CERN $e^+ e^-$ collider
LEP2 \cite{LEPb}, Collider Detectors D0 and CDF at Fermilab
\cite{cdf}); early bounds from Higgs searches at the
LHC\cite{cmshiggs,baglio}; measurements of the $b \rightarrow s +
\gamma$ decay process \cite{bsgamma}: 2.89 $\leq B(b \rightarrow s +
\gamma) \cdot 10^{4} \leq$ 4.21 is employed here (this interval is
larger by 25\% with respect to the experimental determination
\cite{bsgamma} in order to take into account theoretical uncertainties
in the supersymmetric (SUSY) contributions \cite{bsgamma_theorySUSY}
to the branching ratio of the process (for the SM calculation, we
employ the recent NNLO results from Ref.  \cite{bsgamma_theorySM}));
the measurements of the muon anomalous magnetic moment $a_\mu \equiv
(g_{\mu} - 2)/2$: for the deviation, $\Delta a_{\mu} \equiv
a_{\mu}^{\rm exp} - a_{\mu}^{\rm the}$, of the experimental world
average from the theoretical evaluation within the SM we use here the
(2 $\sigma$) range $31 \leq \Delta a_{\mu} \cdot 10^{11} \leq 479 $,
derived from the latest experimental \cite{bennet} and theoretical
\cite{davier} data (the supersymmetric contributions to the muon
anomalous magnetic moment within the MSSM are evaluated here by using
the formulae in Ref. \cite{moroi}); the upper bound on the branching
ratio $BR(B_s^{0} \rightarrow \mu^{-} + \mu^{+})$
\cite{new_Bs,bsmumu}: we take $BR(B_s^{0} \rightarrow \mu^{-} +
\mu^{+}) < 1.5 \cdot 10^{-8}$; the constraints related to $\Delta
M_{B,s} \equiv M_{B_s} - M_{\bar{B}_s}$ \cite{buras_delta,isidori};
the measurements of the decays $B \rightarrow \tau \nu$~\cite{btaunu}
and $R(D) \equiv BR(B \rightarrow D \tau \nu)/BR(B \rightarrow D e
\nu)$~\cite{bdtaunu} (in particular, the compatibility of very light
neutralino masses with the latter four constraints is discussed in
detail in Ref.~\cite{noi_discussing}).

The linear superpositions of bino $\tilde B$, wino $\tilde W^{(3)}$
and of the two Higgsino states $\tilde H_1^{\circ}$, $\tilde
H_2^{\circ}$ which define the four neutralino states, $\chi_i$ (i = 1,
2, 3, 4) are written here as:
\begin{equation}
\chi_i \equiv a_1^{(i)} \tilde B + a_2^{(i)} \tilde W^{(3)} +
a_3^{(i)} \tilde H_1^{\circ} + a_4^{(i)}  \tilde H_2^{\circ}.
\label{neutralino}
\end{equation}

\noindent
The properties of these states have been investigated in detail,
analytically and numerically, in Ref.~\cite{lhc1} for the case when
the smallest mass eigenstate $\chi_1$ (or $\chi$ in short) is light,
{\it i.e.} $m_{\chi} \equiv m_{\chi_1}\lsim 50$ GeV.  Of that analysis
we report here only the main points which are relevant for the present
paper.

We first notice that the lowest value for $m_{\chi}$ occurs when:

\begin{equation}
m_{\chi} \simeq M_1 << |\mu|, M_2.
\label{approx}
\end{equation}

\noindent
since the LEP lower limit on the chargino mass ($m_{{\chi}^{\pm}}
\gsim$ 100 GeV) sets a lower bound on both $|\mu|$ and $M_2$: $|\mu|,
M_2 \gsim$ 100 GeV, whereas $M_1$ is unbound.  Thus, $\chi \equiv
\chi_1$ is mainly a Bino, whose mixings with the other interaction
eigenstates are given by:

 \begin{eqnarray}
\frac{a_2^{(1)}}{a_1^{(1)}} &\simeq& \frac{\xi_1}{M_2} cot_\theta,\nonumber\\
\frac{a_3^{(1)}}{a_1^{(1)}} &\simeq& s_\theta s_\beta \frac{m_Z}{\mu}, \\
\frac{a_3^{(1)}}{a_4^{(1)}} &\simeq&  - \frac{\mu s_\beta}{M_1 s_\beta + \mu c_\beta},\nonumber
\label{approx1}
\end{eqnarray}

\noindent
where $\xi_1 \equiv m_1 - M_1$. These expressions readily follow from
the general analytical formulae given in Ref.~\cite{lhc1} by taking
$\tan \beta \geq$ 10, as consistent with the scenarios discussed
below.

Useful approximate expressions obtain also for the compositions of
the eigenstates corresponding to the asymptotic mass eigenvalues: $m_i
\sim \pm \mu$ and $m_i \sim M_2$. That is:

a) for the  neutralino states $\chi_i$ with $m_i \simeq \pm \mu$,

\begin{eqnarray}
\frac{a_2^{(i)}}{a_1^{(i)}} &\simeq&  \frac{\pm \mu}{M_2  \mp \mu} cot_\theta, 
\nonumber \\
\frac{a_1^{(i)}}{a_3^{(i)}} &\simeq&  \frac{2 \xi_2 s_\theta
(\pm \mu - M_2)}{M_Z  s_\beta ({s_\theta}^2 M_2  \mp \mu)},  \\
\frac{a_3^{(i)}}{a_4^{(i)}} &\simeq& \mp 1 + \frac{\xi_2}{\mu}, 
\nonumber
\label{approx2}
\end{eqnarray}

\noindent
where $\xi_2 \equiv \pm \mu -  m_i$;

b) for the  neutralino state $\chi_i$ with $m_i \simeq M_2$,

\begin{eqnarray}
\frac{a_1^{(i)}}{a_2 ^{(i)}} &\simeq& \frac{\xi_3}{M_2} \tan_\theta, \nonumber \\
\frac{a_1^{(i)}}{a_3^{(i)}} &\simeq& \frac{\xi_3 s_\theta (M_2^2 - \mu^2)}{M_Z (M_2 c_{\beta} +
\mu s_{\beta}) c_{\theta}^2 M_2}, \\
\frac{a_3^{(i)}}{a_4^{(i)}} &\simeq&  - \frac{\mu s_\beta + M_2 c_\beta}{M_2 s_\beta + \mu c_{\beta}}, \nonumber
\label{approx3}
\end{eqnarray}

\noindent
where $\xi_3 \equiv  M_2 - m_i$.

From the above expressions the following relevant properties hold: (i)
$\chi_1$ is mainly a B-ino whose mixing with $\tilde H_1^{\circ}$ is
sizable at small $\mu$, (ii) $\chi_3$ has a mass $|m_3| \simeq |\mu|$
with a large $\tilde H_1^{\circ}-\tilde H_2^{\circ}$ mixing,
independently of $M_2$, (iii) $\chi_2$ and $\chi_4$ interchange their
main structures depending on the value of the ratio $|\mu|/M_2$:
$\chi_2$ is dominantly a W-ino (with a sizable subdominance of $\tilde
H_1^{\circ}$) for $M_2 << |\mu|$ and a maximal $\tilde
H_1^{\circ}-\tilde H_2^{\circ}$ admixture for $M_2 >> |\mu|$, whereas
$\chi_4$ is a maximal $\tilde H_1^{\circ}-\tilde H_2^{\circ}$
admixture for $M_2 << |\mu|$ and a very pure W-ino for $M_2 >> |\mu|$.

For specific spectroscopic schemes, characterized by various internal
hierarchies, we will use the denominations already introduced in
Ref.~\cite{lhc1}, {\it i.e.}: (i) normal hierarchical scheme when $M_2
<|\mu|$, (ii) degenerate scheme when $M_2 \sim |\mu|$, (iii) inverted
hierarchical scheme when $M_2 > |\mu|$ (notice that we always assume
($M_1 << M_2, |\mu|$).

\subsection{Cosmologically inspired scenarios}
\label{sec:cosmo}

If light neutralinos are present in the Universe as relic particles,
their abundance $\Omega_{\chi} h^2$ has to be smaller than the
observed upper bound for cold dark matter (CDM), {\it i.e.}
$\Omega_{\chi} h^2 \leq (\Omega_{CDM} h^2)_{max} = 0.122$ (this
numerical value represents the 2$\sigma$ upper bound to $(\Omega_{CDM}
h^2)_{max}$ derived from the results of Ref. \cite{wmap}).

This requirement implies a lower limit on the neutralino pair
annihilation cross section $\sigma_{\rm ann}$ through the usual
expression:

\begin{equation}
  \Omega_{\chi} h^2 = \frac{x_f}{{g_{\star}(x_f)}^{1/2}} \frac{3.3 \cdot
    10^{-38} \; {\rm cm}^2}{\widetilde{<\sigma_{ann} v>}},
\label{omega}
\end{equation}

\noindent
where $\widetilde{<\sigma_{ann} v>} \equiv x_f {\langle \sigma_{\rm
ann} \; v\rangle_{\rm int}}$, ${\langle \sigma_{\rm ann} \;
v\rangle_{\rm int}}$ being the integral from the present temperature up to
the freeze-out temperature $T_f$ of the thermally averaged product of
the annihilation cross-section times the relative velocity of a pair
of neutralinos, $x_f$ is defined as $x_f \equiv m_{\chi}/T_f$ and
${g_{\star}(x_f)}$ denotes the relativistic degrees of freedom of the
thermodynamic
bath at $x_f$.

The lower bound on $\sigma_{\rm ann}$, implied by the cosmological
upper limit on CDM, combined with the constraints due to accelerator
data and other precision measurements, restricts markedly the overall size of the
supersymmetric parameter space, as depicted in Figs. 1-2 of the second
paper of Ref.~\cite{lowneu}. In particular, it is instrumental in
placing the fore mentioned limit $m_{\chi} \sim$ 9 GeV.

We thus arrive at the formulation of two specific physical scenarios
for the case of light neutralinos of cosmological interest, as
delineated in Ref.~\cite{lhc1}.  These scenarios are determined by the
different ranges of the mass $m_A$ of the CP-odd neutral Higgs boson,
and are summarized in Table~\ref{table:scenarios}.

We have a {\bf Scenario {$\mathcal{A}$}} when 90 GeV $\leq m_A \lsim$
(200-300) GeV (we recall that $m_A \geq$90 GeV is the LEP lower
bound). When $m_A$ is in this range, then the neutralino mass can be
as small as $\sim$ 9 GeV, since the cosmological upper bound is
satisfied due to a sizable contribution to the neutralino pair
annihilation cross section by the exchange of the A Higgs boson in the
{s} channel. For this to be so, the B-ino component of the $\chi_1$
configuration must be maximally mixed with the $\tilde H_1^{\circ}$
component ({\it i. e.}  $a_3^{(1)}/a_1^{(1)} \simeq 0.4)$. From the
second expression in Eq. (\ref{approx1}) one sees that this condition
is satisfied when $\mu$ is small ($|\mu| \sim$ 100-200 GeV).
Moreover, it turns out that $\tan \beta$ must be large ($\tan \beta
\sim$ 30 - 45).  The trilinear coupling is only mildly constrained to
stay in the interval $-1 \lsim A \lsim +1$; the slepton soft mass
$m_{\tilde{l}}$ and the squark soft mass $m_{\tilde q}$ are
unconstrained.  In this scenario, the following hierarchy holds for
the coefficients $a^{(1)}_i$ of $\chi_1$:

\begin{equation}
|a_1^{(1)}| > |a_3^{(1)}| >> |a_2^{(1)}|, |a_4  ^{(1)}|,
\label{hierarchy1}
\end{equation}

\noindent
as easily derivable from Eqs. (\ref{approx1}).

When $m_A \gsim$ (200-300) GeV, the cosmological lower bound on
$\sigma_{\rm ann}$ can be satisfied by a pair annihilation process
which proceeds through an efficient stau-exchange contribution (in the
{\it t, u} channels). This requires that: (i) the stau mass
$m_{\tilde{\tau}}$ is sufficiently light, $m_{\tilde{\tau}} \sim$ 90
GeV (notice that the current experimental limit is $m_{\tilde{\tau}}
\sim$ 87 GeV) and (ii) $\chi_1$ is a very pure B-ino ({\it i.e.} $(1 -
a^{(1)}_1)$ = O($10^{-3}$).  If this is the case, then light
neutralinos can exist, but with a mass above $\sim$ 15-18 GeV
\cite{lowneu,hooper}.  As discussed in Ref.~\cite{lhc1}, conditions
(i) and (ii) require that $|\mu| \gsim$ 500 GeV, $\tan \beta \lsim$
20; $m_{\tilde{l}} \gsim$ (100 - 200) GeV; the parameter $A$ is
typically in the range $-2.5 \lsim A \lsim +2.5$, the other
supersymmetric parameters are not {\it a priori} fixed.  The sector of
the supersymmetric parameter space characterized by these features is
denoted as {\bf Scenario $\mathcal{B}$}. Within this scenario it
follows from Eqs.~(\ref{approx1}) that the following hierarchy holds
for the coefficients $a^{(1)}_i$ of $\chi_1$:

\begin{equation}
|a_1^{(1)}| >> |a_3^{(1)}|, |a_2^{(1)}|, |a_4  ^{(1)}|.
\label{hierarchy2}
\end{equation}

Table I summarizes the representative features of scenarios
$\mathcal{A}$ and $\mathcal{B}$ to be used below for the definition of
our benchmark.

In the present paper we focus our investigation of scenario
$\mathcal{A}$, postponing the discussion of scenario $\mathcal{B}$ for
a subsequent publication.

\begin{table}[t]
\begin{center}
{\begin{tabular}{@{}|c|c|c|c|c|c|@{}}
\hline
~{\rm scenario}~ &  ~$M_1$[GeV]  & $|\mu|$~[GeV]  & $\tan\beta$   &  $m_A$~[GeV] &
$m_{\tilde{l}}~[\rm GeV]$
\\
\hline
\hline
{$\mathcal{A}$} &  $\sim$ 10-14  & 110--140 & 30--45 & $\sim$ 90-110  & --
\\
{$\mathcal{B}$} &  $\sim$ 25  & $\gsim$ 500 & $\lsim$ 20  & $\gsim$
200   & 100--200 \\
\hline
\end{tabular}}
\caption{Representative features for scenarios $\mathcal{A}$ and
$\mathcal{B}$ described in Section
\protect\ref{sec:cosmo}. In scenario $\mathcal{A}$: -1 $\lsim A
\lsim$ +1,
in scenario $\mathcal{B}$: -2 $\lsim A \lsim$ +2.
\label{table:scenarios}}
\end{center}
\end{table}

\section{Signals at the LHC}
\label{sec:signals}

\begin{figure}
\includegraphics[width=0.35\textwidth]{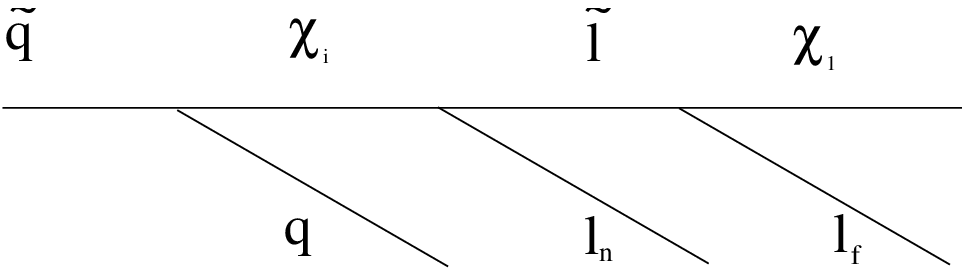}
\,\,\,\,\,\,\,\,\,\,\,\,\,\,\,\,\,\,\,\,\,\,\,\,\,\,\,\,\,\,\,\,\,
\includegraphics[width=0.3\textwidth]{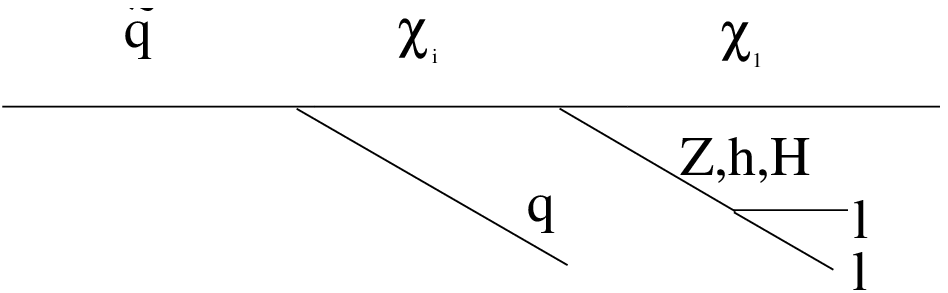}
\caption{Topologies of the decay $\tilde{q}\rightarrow q \chi \bar{l}l$.
Left: sequential decay. Right: branched decay.
\label{fig:decay_topologies}}
\end{figure}

Squarks and gluinos are expected to be copiously produced in the $pp$
scattering processes at the LHC: $pp\rightarrow
\tilde{q}\tilde{q},\tilde{q}\tilde{q}^*,\tilde{g}\tilde{g},
\tilde{q}\tilde{g}$. In turn, squarks, produced either directly or
through gluinos, can generate the sequential decay chains:

\begin{equation}
\tilde{q}\rightarrow q \chi_{i}\rightarrow q \tilde{f}f\rightarrow q
\bar{f}f \chi_1,
\label{eq:sequential}
\end{equation}

\noindent
and the branched ones:

\begin{equation}
\tilde{q}\rightarrow q \chi_{i}\rightarrow q (Z,h,H,A) \chi_1\rightarrow q
\bar{f}f \chi_1,
\label{eq:branched}
\end{equation}

\noindent
where $f$ stands for a fermion, $\bar{f}$ for its supersymmetric
partner; from now on the neutralino subscript $i$ can only take the
values 2, 3 or 4. These two topologies are shown in
Fig.\ref{fig:decay_topologies}.

These are the key processes to be studied at the LHC to measure the
sfermions and neutralinos masses
\cite{atltdr,cmstdr2,Bachacou:2000zb,Allanach:2000kt,su4,sps1a}.  They
would be characterized by hard jets, specific two-body decays and a
transverse missing energy (under the hypothesis of R-parity
conservation). The determination of the masses cannot proceed through
a full reconstruction of the decay chains, since the LSP neutralino
escapes detection, but rather by measurements of specific features in
unidimensional and multidimensional distributions in kinematical
variables.

Typical strategies for determining the sfermion and neutralino masses
consist in: a)~measurements of endpoints in single invariant mass
distributions
\cite{atltdr,cmstdr2,Bachacou:2000zb,Allanach:2000kt,su4,sps1a},
b)~correlations among different invariant mass distributions
\cite{burns}.

\subsection{General properties of the decay chains}
\label{sec:properties}

The decay chains (\ref{eq:sequential})--(\ref{eq:branched}) have in
common the first step, {\it i.e.}  the squark decay
$\tilde{q}\rightarrow q \chi_i$, which can proceed either through
gauge coupling (which involve the gaugino components of $\chi_i$), or
Yukawa coupling (which involve the higgsino components of $\chi_i$).
In the following we will assume for simplicity a situation where the
gluino is decoupled, by taking $M_3\gg m_{\tilde{q}}$. In this case
squarks can be only produced with the same flavor of the partons
inside the protons which induce the hadronic processes at the LHC. As
a consequence of this Yukawa couplings have a subdominant role as
compared to the gauge couplings, since the relative importance of the
Yukawa couplings to the gauge ones depends on the ratio $m_q/m_Z$
($m_q$ and $m_Z$ being the quark mass and the Z-boson mass,
respectively) and heavy flavors are scarce in the proton
composition. In particular, in this case in the process
$\tilde{q}\rightarrow q \chi_i$, the $\chi_i$'s having a dominant
gaugino composition are preferentially produced.

Sequential chains are differentiated from the branched chains by the
features of the decay process undertaken by the intermediate
neutralino state $\chi_i$. In the sequential chain the decay proceeds
through the process: $\chi_i\rightarrow \tilde{f}f\rightarrow \bar{f}f
\chi_1$ with a branching ratio BR($\chi_i\rightarrow
\tilde{f}f\rightarrow \bar{f}f \chi_1$) = BR($\chi_i\rightarrow
\tilde{f}f$) BR($\tilde{f}\rightarrow f \chi_1$). In the following we
will limit our considerations to the most interesting cases, where $f$
is a charged lepton ({\it i.e.} $f = l = e, \mu, \tau$).  The size of
BR($\chi_i\rightarrow \tilde{l}l$) depends sensitively on the $\chi_i$
composition.  If $\chi_i$ is dominantly a gaugino, because of the
universality of the gaugino couplings, the branching ratios
BR($\chi_i\rightarrow \tilde{l}l$) for the three lepton flavours are
about the same ; if $\chi_i$ is dominantly a Higgsino, $\chi_i$ decays
predominantly into a $\tilde{\tau} \tau$ pair.

In the branched chain, $\chi_i$ decays either through the Z-boson or
through a Higgs boson. The first case, {\it i. e.}  $\chi_i
\rightarrow Z + \chi_1$, involves only the Higgsino components of the
two neutralino states; the $Z$ boson subsequently decays into all
(kinematically possible) $\bar{f} f$ pairs according to the Standard
Model branching fractions. The second case, {\it i. e.}  $\chi_i
\rightarrow (h,A,H) + \chi_1$, in order to have a sizable BR, requires
that one neutralino state is dominantly a gaugino, the other
dominantly a Higgsino.  Since in the scenarios considered in the
present paper $\chi_1$ is dominantly a B-ino state, $\chi_i
\rightarrow (h,A,H) + \chi_1$ is of interest when $\chi_i$ is
dominated by the Higgsino components. Because of the hierarchical
character of the Yukawa coupling, the subsequent decays of the Higgs
bosons are dominated by the production of a $b$ -- $\bar{b}$ pair.

A detailed discussion of the branching ratios for the various processes
involved in the decay chains (\ref{eq:sequential})--(\ref{eq:branched})
for the LNM are given in Ref.~\cite{lhc1}.

\section{Selecting a benchmark in the LNM scenario}
\label{sec:selecting_benchmarks}

In the present paper we wish to address the following two points: i)
Is the LNM scenario $\mathcal{A}$ (as defined in Table
\ref{table:scenarios}) detectable at the LHC over the SM background?
ii) Is it possible to kinematically reconstruct the neutralino mass at
the LHC in its low range within the LNM scenario ($m_{\chi}\simeq$ 10
GeV)?

\begin{figure}
\includegraphics[width=0.45\textwidth, bb= 61 63 550 500]{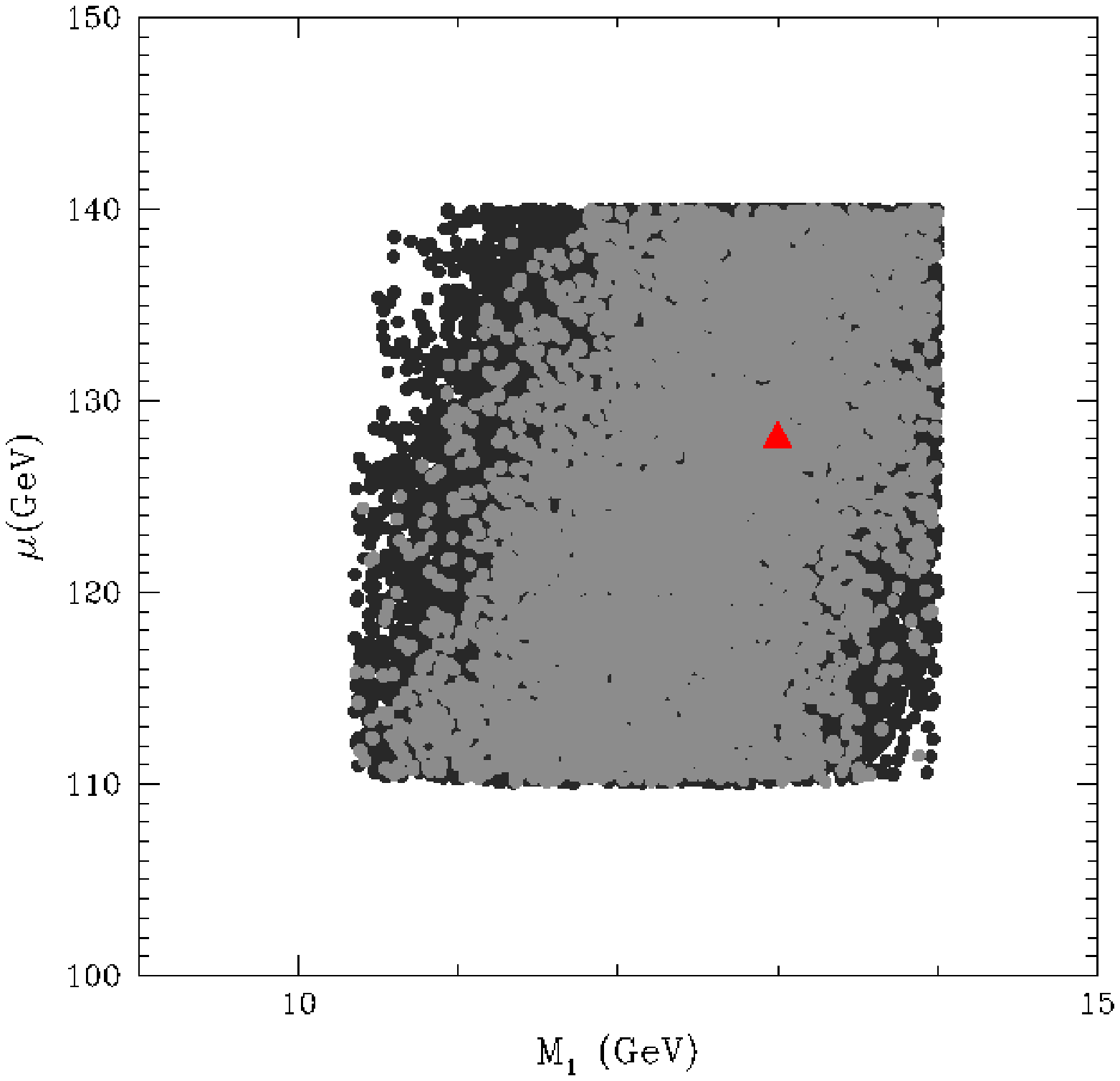} 
\includegraphics[width=0.45\textwidth, bb= 61 63 550 500]{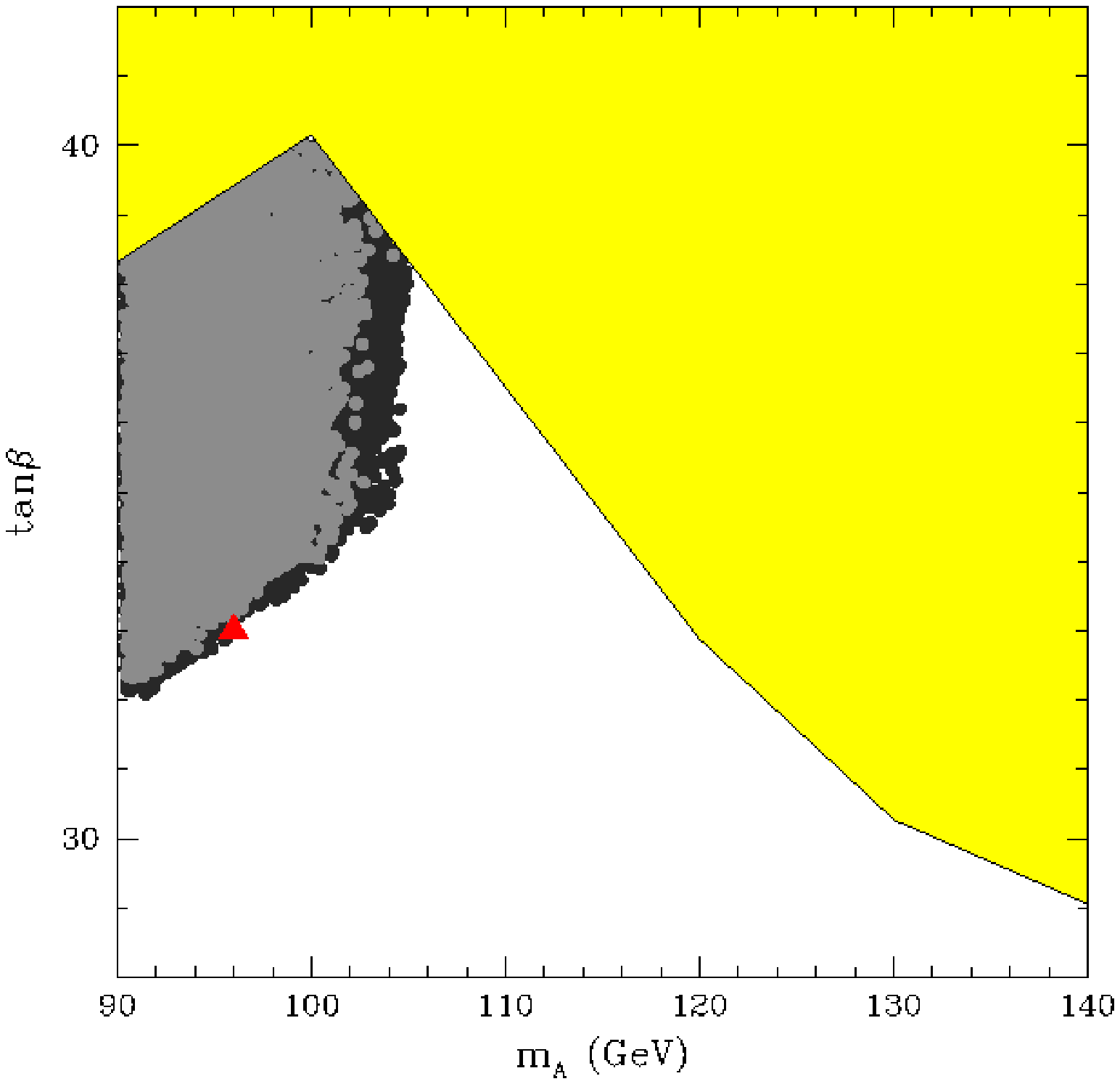} 
\caption{ Scatter plot of the parameters $M_1$ and $\mu$ (left) and of
  the parameters $m_A$ and $\tan\beta$ (right) in the LNM scenario.
  The dots, in black and in gray, correspond to the range of
  parameters given in the text in accord with scenario ${\cal A}$ of
  Table \protect\ref{table:scenarios}; the subset of dots in gray show
  the configurations which fall into region I of
  Fig. \protect\ref{fig:m2_m_sl}, allowing sequential decays through
  production and decay of a $\chi_4$ (see text).  The (red) triangle
  shows the LNM-seq benchmark given in Table \protect
  \ref{table:ln_benchmarks} that is analyzed in detail in the present
  paper. In the right--hand plot the yellow shaded area shows the
  region disallowed in the plane ($m_A$--$\tan \beta$) from the
  results of Refs. \cite{cmshiggs}, as derived in the analysis of
  Ref. \cite{baglio}.
\label{fig:other_parameters}}
\end{figure}

As discussed in Section \ref{sec:cosmo}, one of the basic features of
the LNM scenarios is that some of the SUSY parameters are forced into
rather strict intervals. As shown in Table \ref{table:scenarios}, this
is particularly true in Scenario ${\cal A}$ for the parameters
$M_1,\mu,\tan\beta$ and $m_A$, the latter three parameters being just
beyond the LEP and Tevatron sensitivities.

%

A scatter plot of these 4 parameters where all the experimental
constraints listed in Section \ref{sec:model} are implemented 
is given in Fig.\ref{fig:other_parameters}; here the parameters have
been varied in the following narrow ranges: $10 \, {\rm GeV} \leq M_1
\leq 14 \, {\rm GeV}$, $110 \, {\rm GeV} \leq \mu \leq 140 \, {\rm
  GeV}$, $30 \leq \tan \beta \leq 40$, $90\, {\rm GeV }\leq m_A \leq
105 \, {\rm GeV }$, in accord with the intervals of scenario ${\cal
  A}$ in Table \ref{table:scenarios}. As far as these parameters
are concerned, the choice of a benchmark is quite restricted.

In Fig. \ref{fig:other_parameters} we plot with a triangular symbol
the representative point that we adopt in Table
\ref{table:ln_benchmarks} as our benchmark: $M_1$=14 GeV
(corresponding to $m_{\chi}\simeq$ 11 GeV), $\mu$=126
GeV,$\tan\beta$=34 and $m_A$=97 GeV. The LNM scenario is basically
independent on the remaining six parameters of the model ($M_2$,
$M_3$, $m_{\tilde q}$, $m_{\tilde{t}}$, $m_{\tilde l}$ and $A$), which
are only constrained by the various experimental limits listed in
Section \ref{sec:model}.  In particular LHC physics is very sensitive
through the SUSY production cross section to the $m_{\tilde q}$
parameter (which drives the mass of squarks of the first two families
corresponding to the flavors more abundant in colliding protons) and
to the gluino mass $M_3$.  The LHC early runs have already started to
introduce constraints on these parameters, which, however, strongly
depend on the adopted SUSY scenario
\cite{cms,atlas,atlas2,cms_others,atlas_others}. Following the
approach already adopted in Ref.~\cite{lhc1}, for simplicity in the present
paper we will limit our discussion to the case in which the gluino is
heavier than the squark, and for definiteness we will fix it at the
representative value $M_3$ = 2 TeV.

\begin{figure}
\includegraphics[width=0.45\textwidth ,bb= 44 197 513 631]{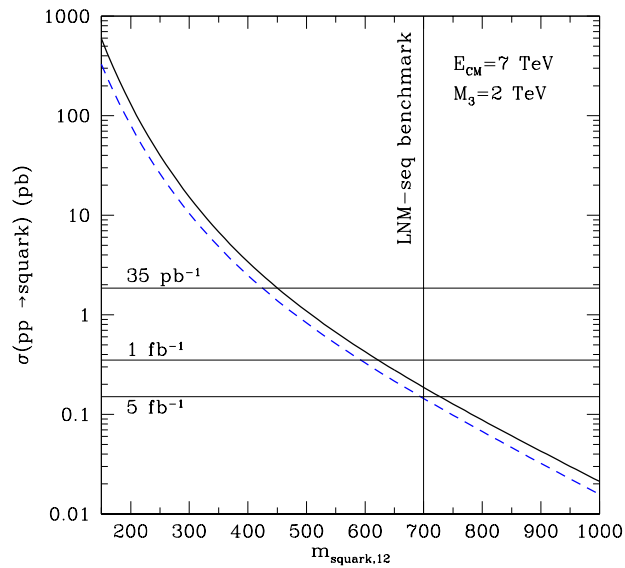} 
\caption{ Squark production cross section at the LHC as a function of
  the mass of the squarks of the first two families
  $m_{squark,12}\simeq m_{\tilde q}$ for a center-of-mass energy
  $E_{CM}$=7 TeV and for a heavy gluino mass, $M_3$=2 TeV. The solid
  line shows the next-to-leading order value calculated with Prospino
  \cite{prospino} while the dashed line shows the same quantity at the
  leading order.  The highest horizontal line marked ``35 pb$^{-1}$''
  shows the bound on the production cross section from early analyses
  of CMS and ATLAS Collaborations at the LHC as calculated in
  Ref.~\protect\cite{noi_impact} for the LNM scenario.  The lower
  horizontal lines show an estimation for the sensitivity of the LHC
  run at $E_{CM}$=7 TeV for two different values of the collected
  luminosity.  The vertical line at $m_{squark}$=700 GeV indicates the
  value of the parameter $m_{\tilde{q}}\simeq m_{squark,12}$ that is
  adopted in the LNM-seq benchmark introduced in Table
  \protect\ref{table:ln_benchmarks}.
\label{fig:cross_section}}
\end{figure}

In Ref. \cite{noi_impact} the possible impact of some early analyses
by the CMS and ATLAS Collaborations at the LHC \cite{cms,atlas,atlas2}
on the LNM scenario was investigated. The data considered there
consisted in the results of searches for supersymmetry in
proton--proton collisions at a center--of--mass energy of 7 TeV with
an integrated luminosity of 35 ${\rm pb}^{-1}$ \cite{cms}, {\it i.e.}
the results of the CMS Collaboration for events with jets and missing
transverse energy \cite{cms}, and those of the ATLAS Collaboration by
studying final states containing jets, missing transverse energy,
either with an isolated lepton (electron or muon) \cite{atlas} or
without final leptons \cite{atlas2}.  As reported in
Refs. \cite{cms,atlas} the data appeared to be consistent with the
expected Standard Model (SM) backgrounds; thus an upper bound on the
SUSY production cross section $\sigma(\rm pp\rightarrow squark)$ at
the LHC for a center-of-mass energy $E_{CM}$=7 TeV was derived in Ref.
\cite{noi_impact} for the LNM scenario. In
Fig. \ref{fig:cross_section} we plot this cross section as a function
of the common mass of the squark of the first two families $m_{squark
  12}\simeq m_{\tilde q}$ for $M_3$=2 TeV. Moreover, the highest
horizontal line marked ``35 pb$^{-1}$'' shows the upper bound on the
same quantity as derived in Ref.~\cite{noi_impact}, that implies a
lower limit $m_{\tilde{q}}\gsim$ 450 GeV when the gluino is heavy. In
the same figure the lower horizontal lines show an estimation for the
sensitivity of the LHC run at $E_{CM}$=7 TeV for two different values
of the collected luminosity, ${\cal L}=$ 1 fb$^{-1}$ and ${\cal L}=$ 5
fb$^{-1}$, naively obtained by scaling down the bound on the cross
section from \cite{noi_impact} with the square root of the
exposition. Assuming ${\cal L}=$ 5 fb$^{-1}$ as the expected total
collected exposition in the LHC run at $E_{CM}$=7 TeV before the stop
scheduled for the end of the year 2011, one can see that a value
$m_{\tilde q}\simeq $700 GeV would be by that time just on the verge
of discovery, possibly already providing a small excess over the
standard model background. For this reason in the following we will
adopt $m_{\tilde q}$=700 GeV as our benchmark value of the soft squark
parameter for the first two families.

\begin{figure}
\includegraphics[width=0.50\textwidth, bb= 44 63 513 531]{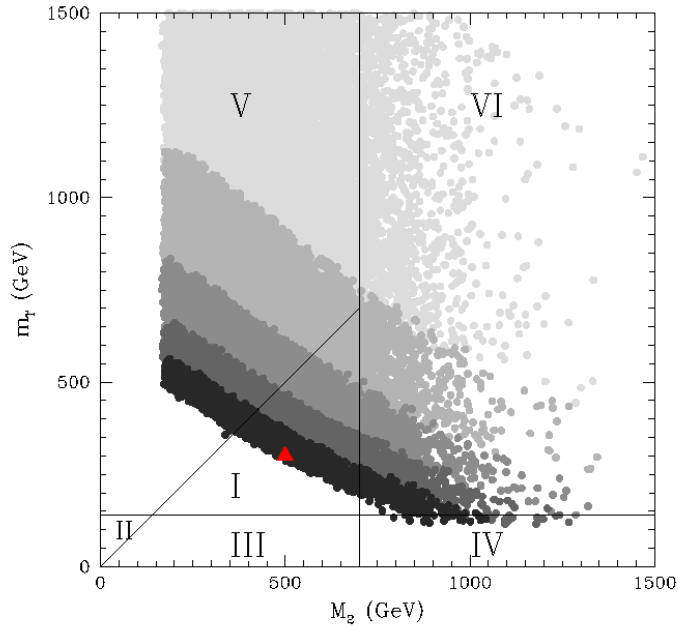} 
\caption{Scatter plot of the parameters $M_2$ and $m_{\tilde{l}}$ in
  the LNM scenario with $m_{\tilde{q}}$=700 GeV and $M_3$=2 TeV. All
  plotted configurations are allowed by the observational constraints
  summarized in Section \protect\ref{sec:model}. In this plot the
  parameters $M_1$, $\mu$, $\tan\beta$ and $m_A$ are sampled in the
  ranges of Scenario ${\cal A}$ indicated in Table
  \protect\ref{table:scenarios}, and their numerical scan is shown in
  Fig. \protect\ref{fig:other_parameters}. The remaining two
  parameters of the model, $A$ and $m_{\tilde{t}}$, are shown in
  Fig.\protect\ref{fig:atrilinear_m_st}. All the configurations are
  subject to the experimental bounds listed in Section
  \protect\ref{sec:model}. The points are plotted in 5 increasingly
  dark tones of gray corresponding to the following sub--intervals for
  the allowed range for the muon anomalous magnetic moment: 31$<\Delta
  a_{\mu}\times 10^{11}<$130, 130$<\Delta a_{\mu}\times 10^{11}<$230,
  230$<\Delta a_{\mu}\times 10^{11}<$330, 330$<\Delta a_{\mu}\times
  10^{11}<$430 and 430$<\Delta a_{\mu}\times 10^{11}<$479.  The
  regions indicated by numbers I, II, III, IV, V and VI correspond to
  different kinematic regimes for the sequential decay (see text for
  details). In particular in regions I,II III and IV sequential decays
  are kinematically accessible through a next-to-lightest neutralino
  $\chi_i$ with $i$=4 (region I), $i$=3,4 (region II), $i$=2,3,4
  (region III) and $i$=2,3 (region IV).  The triangular symbol
  indicates the LNM-seq benchmark given in Table \protect
  \ref{table:ln_benchmarks} that is analyzed in detail in the
  following Sections.
  \label{fig:m2_m_sl}}
\end{figure}

We proceed now to discuss the remaining 4 parameters,
$M_2$,$m_{\tilde{l}}$, $m_{\tilde{t}}$ and $A$. As discussed in
Section \ref{sec:properties} the properties of the sequential decay of
Eq. (\ref{eq:sequential}) that we wish to analyze depend sensitively
on the hierarchy among the masses of the particles involved, i.e. on
the masses of the squarks, of the next-to-lightest neutralino and of
the slepton. In our scenario these three mass scales are driven by
$m_{\tilde q}$, $m_{\tilde{l}}$ and $M_2$ or $\mu$, determining in
particular whether the spectrum of neutralinos is normal or inverted
and if the decay $\chi_{i=2,3,4}\rightarrow \tilde{l} l$ is
kinematically allowed. Since we have fixed $m_{\tilde q}$=700 GeV and
in Scenario ${\cal A}$ the $\mu$ parameter is constrained to the
narrow range 110 GeV$\lsim \mu \lsim$ 140 GeV, in
Fig. \ref{fig:m2_m_sl} we discuss this mass hierarchy in the plane of
the remaining two parameters, $M_2$ and $m_{\tilde{l}}$. In this plane
we schematically represent with a line at the constant value
$m_{\tilde{l}}$=140 GeV the scale of the $\mu$ parameter.  Then, with
the exception of the narrow band where $M_2\simeq \mu$, one has
$\mu\simeq m_{\chi_{2,3}}$, $M_2\simeq \chi_4$ when $M_2>\mu$ (normal
hierarchy) and $\mu\simeq m_{\chi_{3,4}}$, $M_2\simeq m_{\chi_2}$ when
$M_2>mu$ (inverted hierarchy). Moreover $m_{\tilde{l}}$ fixes the
scale of the slepton masses. This implies that schematically one can
divide the $M_2$--$m_{\tilde{l}}$ plane in six regions:

\begin{itemize}

\item Region I ($M_2>m_{\tilde{l}}>\mu$). Here only the decay
  $\chi_{4}\rightarrow \tilde{l}l$ is kinematically allowed, with $m_{\chi_4}\simeq M_2$.

\item Region II ($\mu>m_{\tilde{l}}>M_2$). Here only the decays
  $\chi_{3,4}\rightarrow \tilde{l}l$ are kinematically allowed, with
  $m_{\chi_{3,4}}\simeq \mu$;

\item  Region III ($M_2>\mu>m_{\tilde{l}}$). All decays
          $\chi_{2,3,4}\rightarrow \tilde{l}l$ are kinematically allowed,

        \item Region IV ($M_2>m_{\tilde q}$=700 GeV,
          $\mu>m_{\tilde{l}}$). Here only the decays
  $\chi_{2,3}\rightarrow \tilde{l}l$ are kinematically allowed, with
  $m_{\chi_{2,3}}\simeq \mu$; 
\item Regions V and VI. No sequential decays are kinematically allowed.
\end{itemize}

\begin{figure}
\includegraphics[width=0.50\textwidth, bb= 44 63 513 531]{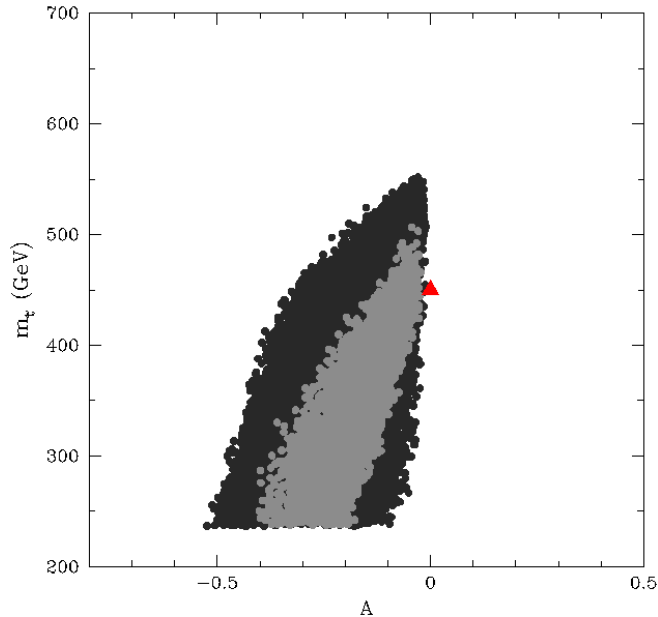} 
\caption{Scatter plot of the parameters $A$ and  $m_{\tilde{t}}$ in the LNM scenario. 
The color code is the same as in Figure \protect\ref{fig:other_parameters}.
The (red) triangle shows the LNM-seq  benchmark given in Table 
\protect \ref{table:ln_benchmarks}.
\label{fig:atrilinear_m_st}}
\end{figure}

In the same figure the scatter plot represents a scan of the LNM
parameter space with 110 GeV$<M_2<$ 1500 GeV, 110 GeV
$<m_{\tilde{l}}<$ 1500 GeV, while the other parameters are in the
ranges of Scenario ${\cal A}$ given in Table \ref{table:scenarios}.
The points are plotted in 5 increasingly dark tones of gray
corresponding to the following sub--intervals for the allowed range
for the muon anomalous magnetic moment: 31$<\Delta a_{\mu}\times
10^{11}<$130, 130$<\Delta a_{\mu}\times 10^{11}<$230, 230$<\Delta
a_{\mu}\times 10^{11}<$330, 330$<\Delta a_{\mu}\times 10^{11}<$430 and
430$<\Delta a_{\mu}\times 10^{11}<$480. In this way it is possible to
see that the kinematic regions II and III (at least for the particular
choice of $m_{\tilde q}$ adopted here) are not allowed by the upper
bound on $\Delta a_{\mu}$\footnote{The quantity $\Delta a_{\mu}$ does
  not depend on the SUSY hadronic sector, so a change in the
  $m_{\tilde{q}}$ parameter would imply only a shift of the vertical
  line separating kinematic regions I, III and V from regions IV and
  VI, without modifying the scatter plot. This implies that for
  \msq$\gsim$ 800 GeV the kinematic region III would be allowed for
  sequential decays}. This restricts the present discussion to the
possibility of having a sequential decays only in regions I and IV.
In region IV sequential decays proceed through production and decays
of very light next-to-lightest neutralinos $\chi_{2,3}$ with
$m_{\tilde{l}}\lsim m_{\chi_{12}}\simeq\mu\lsim 140$ GeV that,
according to the discussion of Section \ref{sec:model}, are of
higgsino type. These features make the detection of sequential decays
in this case quite challenging, since the corresponding branching
ratio is suppressed both by the small available phase space, and by
the fact that, due to their higgsino nature, a large fraction of the
$\chi_{2,3}$ particles decay through the branched topology to a $Z$ or
a Higgs boson (see Fig. \ref{fig:decay_topologies}, right). Moreover,
the higgsino nature of the $\chi_{2,3}$ particles also implies that
when a sequential decay actually takes place it mainly proceeds to tau
final states that are more difficult to measure compared to muons and
electrons.  As a consequence of this, for the choice of a benchmark
for our discussion of sequential decays we decide to focus on the only
remaining possibility, i.e. region I. In
Figs. \ref{fig:other_parameters} and \ref{fig:atrilinear_m_st} the
configurations belonging to region 1 are plotted in grey.

Region I looks more promising than region IV for sequential decays. In
fact in this case decays proceed through a $\chi_4$ with
$m_{\chi_4}\simeq M_2$ which is of Wino type implying a smaller
coupling to the $Z$ and Higgs bosons which reduces branched decays and
leading to comparable signals to electrons, muons and taus in
sequential ones. Moreover, depending on the choice of the $M_2$ and
$m_{\tilde{l}}$ parameters, the phase space available to both decays
$\tilde{q}\rightarrow q \chi_4$ and $\chi_4\rightarrow l \tilde{l}$,
can be sizeable. As the scatter plot of Figure \ref{fig:m2_m_sl}
shows, this still allows for a wide range of possibilities. With the
spirit of choosing light values for both the $\chi_4$ and the sleptons
masses and to maximize at the same time the phase space available to
the decay, in Table \ref{table:ln_benchmarks} we choose as our
benchmark the values $M_2$=500 GeV and $m_{\tilde{l}}$=300 GeV,
corresponding to the point shown in Fig. \ref{fig:m2_m_sl} with a
triangular symbol lying somewhat in the center of the triangle of
region I and close to the lower bound on both parameters from the muon
anomalous magnetic moment.

We conclude the present discussion with the last two parameters, $A$
and $m_{\tilde{t}}$. Apparently, since they do not affect directly
either the neutralino relic abundance in the LNM scenario or the
sequential decays we wish to discuss, pinning down a value for these
two parameters may seem quite arbitrary. However, as discussed in
Ref.~\cite{noi_impact}, the combination of the experimental
constraints from the $b\rightarrow s \gamma$ and the $B \rightarrow
\tau \nu$ decays may induce a strong correlation between A and
$m_{\tilde{t}}$, restricting their range of variation. This is shown
in Fig.\ref{fig:atrilinear_m_st}, where a scatter plot of these two
parameters is given and both constraints are applied. Notice that the
correlation is further enhanced if configurations plotted in grey are
considered, corresponding to points falling in region I of
Fig.\ref{fig:m2_m_sl}.  As pointed out in
Ref.~\protect\cite{noi_discussing}, constraints from rare B--meson
decays are affected by uncertainties both in experimental measurements
and theoretical estimates, so should be considered with care. In
particular when the bound from the $B\rightarrow \tau \nu$ is not
implemented the correlation between $A$ and $m_{\tilde{t}}$ is no
longer present \cite{noi_impact}. With this caveat, and for the sake
of definiteness, we may choose our benchmark as a configuration within
the grey region of Fig. \ref{fig:atrilinear_m_st}. In particular, in
order to kinematically suppress the decay $\chi_4\rightarrow t
\tilde{t}$ and maximize the leptonic sequential signature, we choose
as our benchmark a configuration with a value of $m_{\tilde{t}}$ close
to the upper edge of the allowed range. Our final choice for the last
two benchmark parameters is plotted in Fig.\ref{fig:atrilinear_m_st}
with the triangular symbol, and corresponds to the values given in
Table \ref{table:ln_benchmarks}: $A$=-0.08 and $m_{\tilde{t}}$=444 GeV.

\begin{table}[t]
\begin{center}
{\begin{tabular}{@{}|c|c|c|c|c|c|c|c|c|c|c|@{}}
\hline
~{\rm benchmark}~ &  ~$M_1$[GeV]  & ~$M_2$[GeV] & ~$M_3$[GeV] & 
$\mu$~[GeV]  & $\tan\beta$   &  $m_A$~[GeV] &
$m_{\tilde{l}}~[\rm GeV]$ &  $m_{\tilde{q}}~[\rm GeV]$&  $m_{\tilde{t}}~[\rm GeV]$ & $A$ \\
\hline
\hline
{LNM-seq} &  14  & 500 & 2000 & 126 & 34  & 97
& 300 & 700 & 444 & -0.08
\\
\hline
\end{tabular}}
\caption{The LNM-seq benchmark analyzed in the present paper. 
\label{table:ln_benchmarks}}
\end{center}
\end{table}

\section{Early discovery of light neutralinos at the LHC}
\label{seq:early}

The LHC has already started to put bounds on the supersymmetric
parameter space. In particular, the very constrained SUGRA scenario,
in which soft masses and the trilinear coupling are all unified at the
GUT scale and the $\mu$ and $m_A$ parameters are predicted by
radiative elecroweak symmetry breaking, appears already to be
disfavored by the data
\cite{cms,atlas,atlas2,cms_others,atlas_others}. Actually, in this
scenario lower bounds on both gluino and squark masses are already
close to the TeV range, in tension with the naturalness picture that
is considered one of the motivations of SUSY in the first place.

As already mentioned in the previous section, in order to discuss LHC
bounds in the LNM scenario, a dedicated analysis was performed in
Ref.~\cite{noi_impact} for some of the specific signatures searched by
ATLAS and CMS (namely jets+missing transverse energy and one isolated
lepton or jets +missing transverse energy and no leptons). As
discussed in that paper, if squark soft masses of the three families
are assumed to be degenerate, the combination of the ensuing LHC
constraints on squark and gluino masses with the experimental limit on
the $b \rightarrow s + \gamma$ decay imply a lower bound on the
neutralino mass $m_{\chi}$ that can reach the value of 11.9 GeV when
the gluino mass is at its lower bound, but is essentially unchanged
for a heavy gluino. However this bound on $m_{\chi}$ is no longer in
place when, as in the present analysis, the universality condition
among squark soft parameters is relaxed. This implies that for
non-universal squark masses the lower bound on the neutralino mass
remains at the value 9 GeV mentioned in the Introduction.
  
The LHC is expected to collect ${\cal L}\simeq$5 fb$^{-1}$ of
integrated luminosity at the end of the 2011 run at a center-of-mass
energy of 7 TeV. In order to estimate the expected signal at the end
of the 7 TeV run of the LHC for the LNM-seq benchmark introduced in
the previous Section we have used ISAJET \cite{isajet}, applying the
same kinematic cuts as described in Ref. \cite{cms} for the early
discovery signature of jets +missing transverse energy and no
leptons. The result of the simulation is 260 events from SUSY compared
to 133 events expected from backgrounds.  The background estimation is
an extrapolation based on CMS measurements.  The expected number of
SUSY events $N$ is related to the SUSY production cross section
$\sigma\equiv \sigma(\rm pp\rightarrow squark)$ and to the luminosity
${\cal L}$ by the relation $N=\epsilon\times {\cal L}\times \sigma$,
where $\epsilon$ is the total efficiency due to selection cuts, that
for the LNM-seq benchmark we estimate $\epsilon$=0.21. So, in
agreement to the discussion of the previous Section, the LNM-seq
benchmark is expected to provide a slight excess over the background,
namely at the level of a $\simeq$ 3.2 $\sigma$ significance assuming
that our estimation on the background has a 5\% relative uncertainty.

Such an early hint of SUSY in the 7 TeV run of the LHC would not,
however, allow to draw any conclusions on the mass and properties of
the neutralino, let alone whether the observed excess is compatible to
an LNM scenario or not. In fact, since the neutralino escapes
undetected, its mass can only be reconstructed by observing the
sequential decay introduced in Eq. (\ref{eq:sequential}) where the
observed fermions are either muons or electrons in order to have a
better discrimination of the signal over the hadronic background.  In
the corresponding final state of 2 jets+missing transverse energy and
2 isolated leptons we estimate from the above simulation $\simeq$2
signal events at ${\cal L}$=5 fb$^{-1}$ (corresponding to an
efficiency $\epsilon$=0.00155) obviously insufficient to get any
information about the masses. For this kind of analysis $E_{cm}$=14
TeV and a higher collected luminosity will be needed.

\section{LNM spectroscopy at the LHC with $E_{CM}$=14 TeV}
\label{sec:spectroscopy}


The use of kinematic endpoints to reconstruct the mass spectrum in a
sequential decay chain where the lightest particle escapes detection
and with the topology as shown in Fig.\ref{fig:decay_topologies} has
been widely discussed in the
literature\cite{endpoints_others,Allanach:2000kt,sps1a,burns}. This
technique is based on the simple idea of reconstructing the four
unknown masses of the problem
$(m_{\chi},m_{\tilde{l}},m_{\chi_{i}},m_{\tilde{q}})$ by inverting the
four observable kinematic endpoints
$(m_{\ell\ell}^{max},m_{j\ell\ell}^{max},m_{j\ell(lo)}^{max},m_{j\ell(hi)}^{max})$,
where $j$ indicates the jet produced by the emitted quark and
$m_{j\ell(lo)}\equiv$ Min($m_{j\ell_n}$,$m_{j\ell_f}$), $m_{j\ell(hi)}\equiv$
Max($m_{j\ell_n}$,$m_{j\ell_f}$) are defined in order to remove the
ambiguity between the near and the far leptons $\ell_n$, $\ell_f$ (shown in
Fig.\ref{fig:decay_topologies}) which are not experimentally
distinguishable. The expected value of such endpoints are given in
Table \ref{table:endpoints} for the LNM-seq benchmark introduced in
Section \ref{sec:selecting_benchmarks}. For completeness we give the
analytic expressions of the endpoints as a function of the masses in
Appendix \ref{sec:analytic_endpoints}.


\begin{table}[t]
\begin{center}
\begin{tabular}{|c||c|c||} \hline
Variable  & LNM-seq & LNM-seq$^{\prime}$   \\ 
\hline 
 $m_{\chi}$ &   11  &   263       \\
 $m_{\tilde{l}}$ &    305  &  383  \\
 $m_{\chi_{i}}$ &   515  &  688  \\
 $m_{\tilde{q}}$ &   703  &  896     \\
\hline
\hline
 $m_{\ell\ell}^{max}$ &   \multicolumn{2}{c||}{415 (417.5$\pm$ 3.5)}     \\
 $m_{j\ell\ell}^{max}$ &   \multicolumn{2}{c||}{632 (631.3$\pm$ 3.8)}     \\
 $m_{j\ell(lo)}^{max}$ &   \multicolumn{2}{c||}{338  (342.2  $\pm$4.3) } \\
 $m_{j\ell(hi)}^{max}$ &   \multicolumn{2}{c||}{477 (483$\pm$ 14)} \\
 $m_{j\ell\ell(\theta>\pi/2)}^{min}$ &   \multicolumn{2}{c||}{{400 (399.3 $\pm$ 1.7)}} \\
\hline 
\hline 
 $n^{\prime},p^{\prime}$ &  282,385  & 232,477  \\
\hline
\end{tabular}
\end{center}
\caption{Expected endpoints (in GeV) of the kinematic variables
  $(m_{\ell\ell},m_{j\ell\ell},m_{j\ell(lo)},m_{j\ell(hi)},m_{j\ell\ell(\theta>\pi/2)}^{min})$
  for the LNM-seq benchmark of Table
  \protect\ref{table:ln_benchmarks}, calculated using the expressions
  summarized in Appendix \ref{sec:analytic_endpoints}. In parenthesis
  we give measurements of the same quantities analyzing the output of
  a simulation of proton--proton collisions at $E_{CM}$=14 TeV
  assuming an integrated luminosity of 100 fb$^{-1}$ (see text). The
  benchmark LNM-seq$^{\prime}$ indicates the duplicate model of
  LNM-seq, i.e. a different mass pattern providing the same values of
  the observed endpoints in one-dimensional
  distributions\protect\cite{burns}. In the last line the coordinates
  of the point
  ($m_{j\ell(lo)}$,$m_{j\ell(hi)}$)=($n^{\prime}$,$p^{\prime}$) can
  break the degeneracy between the two duplicated models. For this
  particular mass pattern it is not possible to measure the exact
  values $n^{\prime}$,$p^{\prime}$, since they lie on a straight
  boundary of the two--dimensional distribution (see
  Fig. \protect\ref{fig:mljhi_mljlo}).  However this is sufficient to
  break the degeneracy (see text). For this reason we do not provide a
  measured value of $n^{\prime}$,$p^{\prime}$ from the simulation.
  The analytic expressions of ($n^{\prime}$,$p^{\prime}$) as a
  function of the physical masses are given in Appendix
  \ref{sec:analytic_endpoints}. \label{table:endpoints}}
\end{table}

Two problems however arise in this apparently straightforward
procedure. The first issue is related to the fact that for particular
mass combinations in the sequential decay the four aforementioned
endpoints are not independent, since the following relation
holds\cite{sps1a}:

\be
(m_{j\ell\ell}^{max})^2=(m_{\ell\ell}^{max})^2+(m_{j\ell(hi)}^{max})^2.
\label{eq:correlation}
\ee

In particular, this is true whenever
$m_{\chi}<m_{\tilde{l}}^2/m_{\tilde{q}}$. Taking into account the
experimental constraint $m_{\tilde{l}}\gsim$ 100 GeV implies that for
a light neutralino of mass $m_{\chi}=10$ GeV the correlation
(\ref{eq:correlation}) is verified if $m_{\tilde{q}}\lsim$ 1 TeV and,
in particular, holds in the LNM-seq benchmark. To compensate for the
occurrence of only three independent variables of the physical masses
instead of four an additional measurement is needed, for instance the
{\it lower} kinematic endpoint $m_{j\ell\ell(\theta>\pi/2)}^{min}$
introduced in Ref.~\cite{Allanach:2000kt}, which corresponds to the
lower bound of the $m_{j\ell\ell}$ histogram with the additional
constraint
$(m_{\ell\ell}^{max})^2/2<(m_{\ell\ell})^2<(m_{\ell\ell}^{max})^2$.

The second issue related to the analysis of endpoints is that when
Eq. (\ref{eq:correlation}) is verified the inversion procedure is
known to have multiple solutions, so that the determination of the
masses $(m_{\chi},m_{\tilde{l}},m_{\chi_{i}},m_{\tilde{q}})$ is
non--unique.  This is indeed what happens in the LNM-seq benchmark, as
shown in Table \ref{table:endpoints}. In the top part of the Table we
report the values for the masses
$(m_{\chi},m_{\tilde{l}},m_{\chi_{i}},m_{\tilde{q}})$. In the middle
part of the same Table we give the expected endpoints of the kinematic
variables
$(m_{\ell\ell},m_{j\ell\ell},m_{j\ell(lo)},m_{j\ell(hi)},m_{j\ell\ell(\theta>\pi/2)}^{min})$
in sequential decays, evaluated by using the expressions summarized in
Appendix \ref{sec:analytic_endpoints}. By using mass--inversion
formulas \cite{burns} one finds that starting from the set of the
end--point values displayed in the Table one recovers the input set of
mass values together with a second set of masses. In the Table such
additional mass spectrum, hereafter referred as the duplicate of
LNM-seq, is denoted as LNM-seq$^{\prime}$. As will be discussed in the
following, in order to overcome this duplication problem it is
necessary to go beyond endpoints in one--dimensional histograms and to
analyze the correlations among different invariant masses in
two--dimensional plots \cite{burns}.

We wish now to discuss if the procedure outlined above can be
applicable to determine the mass spectrum of the LNM-seq benchmark
using the LHC data at $E_{CM}$=14 TeV. In order to do this, we
simulate proton-proton collisions at $E_{CM}$=14 TeV using ISAJET and
select events with two jets, two isolated leptons and missing
transverse energy.

Notice that fast detector simulation tools which have been developed
for the study of specific Supersymmetric scenarios such as in
Supergravity--inspired (SUGRA) benchmarks are not available for the
model under consideration here. So, in order to take into account the
detector response, and specifically the uncertainty in the
reconstruction of jet energies, in our simulation we smear the energy
$E$ of quarks and gluons in the final state. In particular, we apply a
resolution which depends on energy as 0.9 $\sqrt{E}$. The Missing
Transverse Energy (MET) is then determined by the vector sum of the
energies of the neutrinos and the LSP plus any smearing applied to the
hadronic jets.  Due to the large mass of squarks and gluinos the
dominant background is expected to be due to $t\bar{t}$ production. In
particular, as shown below, large cut values are needed for an
effective separation between the signal and the background. We apply
to the output of our simulation the following cuts:

\begin{itemize}

\item The two leptons ($e^+e^-$ or $\mu^+\mu^-$) are required to
  satisfy $|\eta|<2.4$ (where $\eta\equiv-\ln[\tan(\theta/2)]$ is the
  pseudo--rapidity and $\theta$ is the angle with
  the beam axis) and $p_T>20$ GeV (where $p_T$ is the
  transverse momentum).

\item The kinematic separation between outgoing states is required to
  be $\Delta R>0.5$, where $\Delta R\equiv \sqrt{(\Delta
    \eta)^2+(\Delta \phi)^2}$ and $\phi$ is the azimuthal angle.

\item A missing transverse energy $E_T>300$ GeV is required, in order
  to indicate the presence of high energy neutralinos.

\item The scalar sum of the transverse momenta $p_T$ of leptons and
  jets is required to be larger than 600 GeV.

\item In the study of 2-dimensional distributions we have removed
  events where the invariant mass of the two outgoing leptons falls in
  the range 87 GeV $<M_{\ell\ell}<$ 97 GeV, in order to subtract $Z$
  boson decays. These events include also those produced in the
  branched decays shown in Fig.\ref{fig:decay_topologies}, right.

\end{itemize}

Events with more than two jets or two leptons are rejected to minimize
the effects of combinatorics. Out of the two jets in the event, only
one must be associated to the dilepton in order to construct the
$m_{j\ell(lo)}$, $m_{j\ell(hi)}$ invariant masses. The jet-dilepton
pairing is found by choosing the combination that yields the smallest
value for the $m_{\ell\ell j}$ invariant mass.

\begin{figure}
\includegraphics[width=0.7\textwidth]{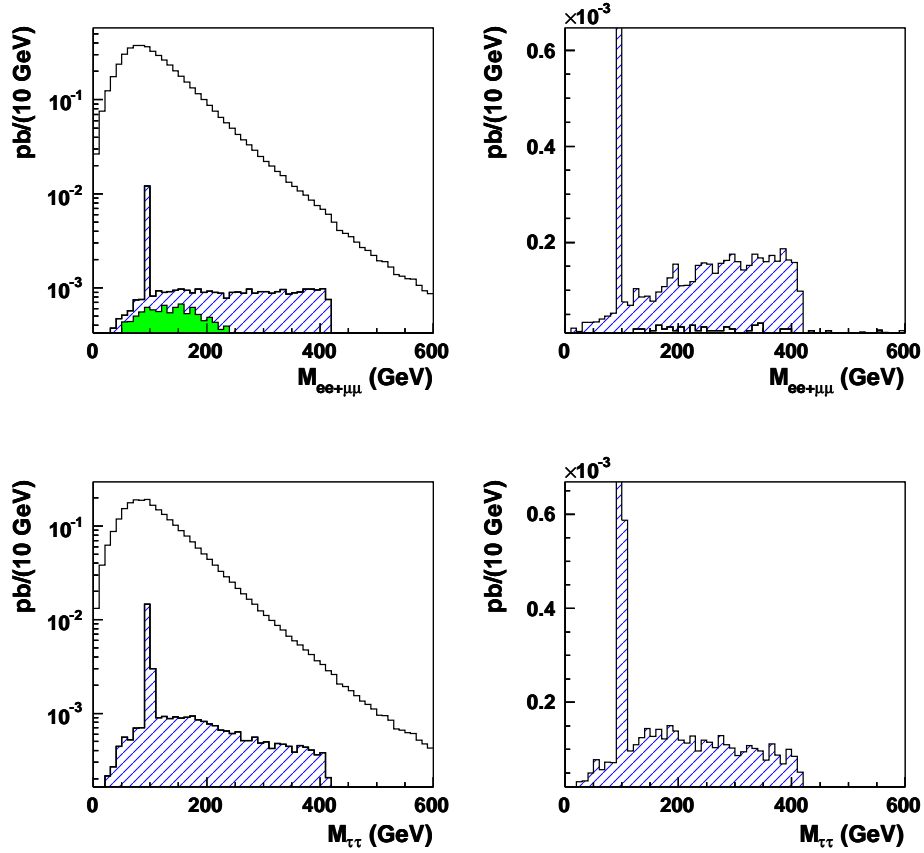}
\caption{Event distribution as a function of the final leptons
  invariant mass $m_{\ell\ell}$ for the benchmark LNM--seq.  Plots on
  the left show the dilepton invariant mass at the preselection level
  while plots on the right show the same histograms after the cuts
  described in Section \ref{sec:spectroscopy} are applied, with the
  exception of the subtraction of the $Z$ peak (which is clearly
  visible). Upper plots show the case when the final leptons are given
  by $e^{+}e^{-}$ and $\mu^{+}\mu^{-}$, while lower plots show the
  $\tau^{+}\tau^{-}$ final state. In the plot on the upper left the
  hatched histogram shows the distribution of the
  $e\bar{e}+\mu\bar{\mu}$ events, while the (green) solid histogram
  shows the same for the $e\bar{\mu}+\mu\bar{e}$ events, which
  provides an estimation of the expected contribution from the SUSY
  background (see text). The shaded histogram in the plot on the upper
  right shows the background--subtracted distribution given by the
  difference between the $e\bar{e}+\mu\bar{\mu}$ and the
  $e\bar{\mu}+\mu\bar{e}$ histograms.  In the case of the lower plots
  the hatched histogram on the left shows the distribution of
  $\tau^{+}\tau^{-}$ events before cuts, while that on the right shows
  the same quantity after cuts. In all plots the white histogram shows
  the $t\bar{t}$ backgrounds.
\label{fig:mll}}
\end{figure}

\begin{figure}
\includegraphics[width=0.7\textwidth]{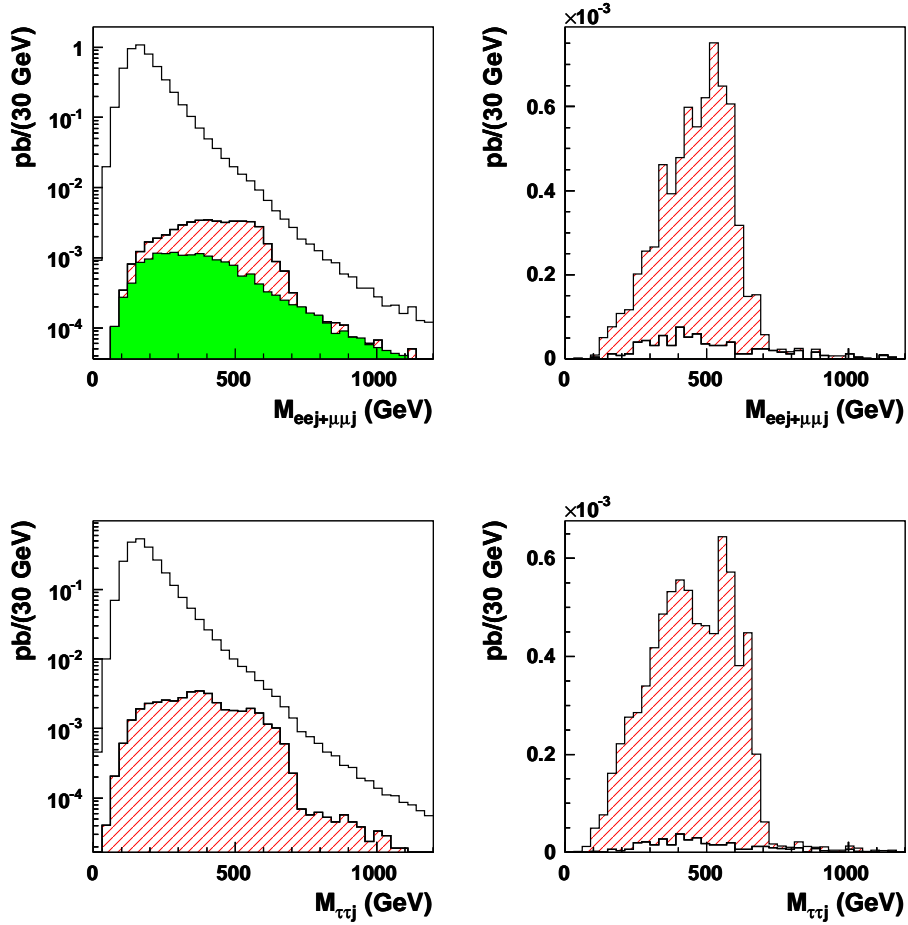}
\caption{Event distribution as a function of the invariant mass
  $m_{\ell\ell j}$ of the final leptons and the jet for the benchmark
  LNM--seq. The color code and the cuts are the same as in
  Fig.\protect\ref{fig:mll}.\label{fig:mllj}}
\end{figure}

The expected distribution of the events with 2 jets+2 leptons+missing
transverse energy produced in the simulation is shown in
Fig.\ref{fig:mll} as a function of the invariant mass $m_{\ell\ell}$
and in Fig.\ref{fig:mllj} as a function of $m_{\ell\ell j}$. In both
Figures the plots on the left show events at the pre--selection level,
while those on the right show the same distributions after the cuts
described above, with the exception of the subtraction of the $Z$
peak. Moreover, upper plots show the case when the final leptons are
electrons or muons, while lower plots show the $\tau^+ \tau^-$ final
state.  In all plots the white histogram shows the $t\bar{t}$
backgrounds, which, as can be seen, is strongly suppressed by the
cuts.

A different and potentially sizeable source of background making the
determination of endpoints difficult is also represented by SUSY
events where the two charged leptons used to calculate the invariant
mass are not originated in a sequential decay, but are produced
instead by the decays of charginos originating from different decay
chains. In the upper--right plots of Figs. \ref{fig:mll} and
\ref{fig:mllj} these undesired events are subtracted exploiting the
fact that in this case the flavors of the two leptons is uncorrelated,
while when the two leptons are produced in the same sequential decay
they have the same flavor. For this reason an effective subtraction of
this background is obtained by taking the difference between the
number of $e\bar{e}$+$\mu\bar{\mu}$ same--flavor events minus the
number of events where the flavor of the final leptons is different,
$e\bar{\mu}$+$\mu\bar{e}$. This is indeed an effective technique to
subtract chargino decays and to allow a better identification of the
endpoints, since the difference of the two distributions is expected
to drop beyond the boundaries of the sequential process. In
particular, in the plots on the upper left of Figs.\ref{fig:mll} and
\ref{fig:mllj} the hatched histogram shows the distribution of the
$e\bar{e}+\mu\bar{\mu}$ events while the solid histogram shows the
same for the $e\bar{\mu}+\mu\bar{e}$ events, which provides an
estimation of the expected contribution from the SUSY
background. Moreover, the shaded histogram in the plots on the upper
right show the background--subtracted distribution given by the
difference between the $e\bar{e}+\mu\bar{\mu}$ and the
$e\bar{\mu}+\mu\bar{e}$ histograms. In the case of the lower plots the
hatched histogram on the left shows the distribution of
$\tau^{+}\tau^{-}$ events before cuts, while that on the right the
same quantity after cuts. Comparison of left--hand figures and
right--hand ones prove the overall effectiveness of the applied cuts
to subtract the standard model background, in particular from top
decays.

\begin{figure}
\subfigure[]{\includegraphics[width=0.4\textwidth]{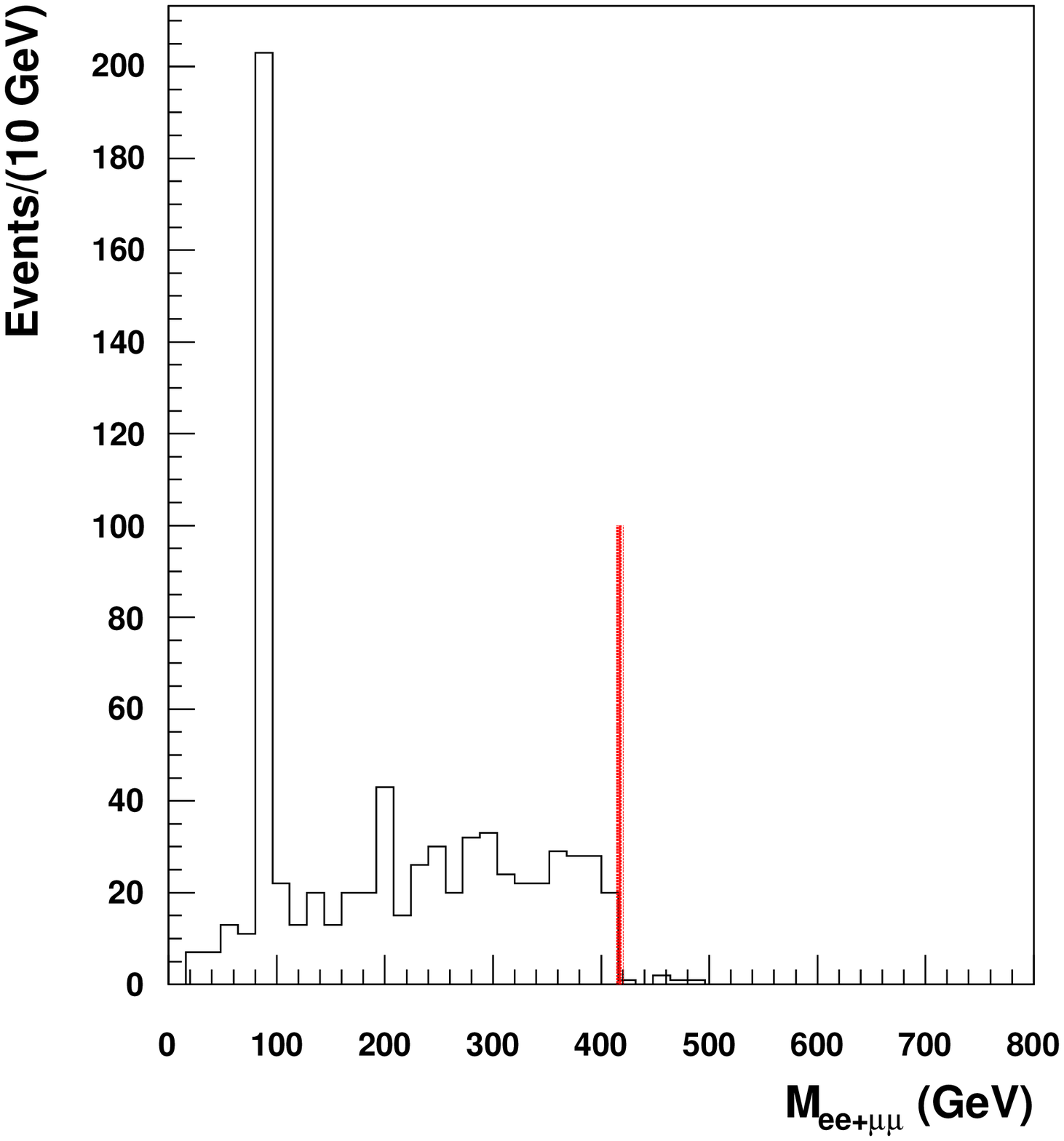} 
\label{fig:mll-100fb}}
\subfigure[]{\includegraphics[width=0.4\textwidth]{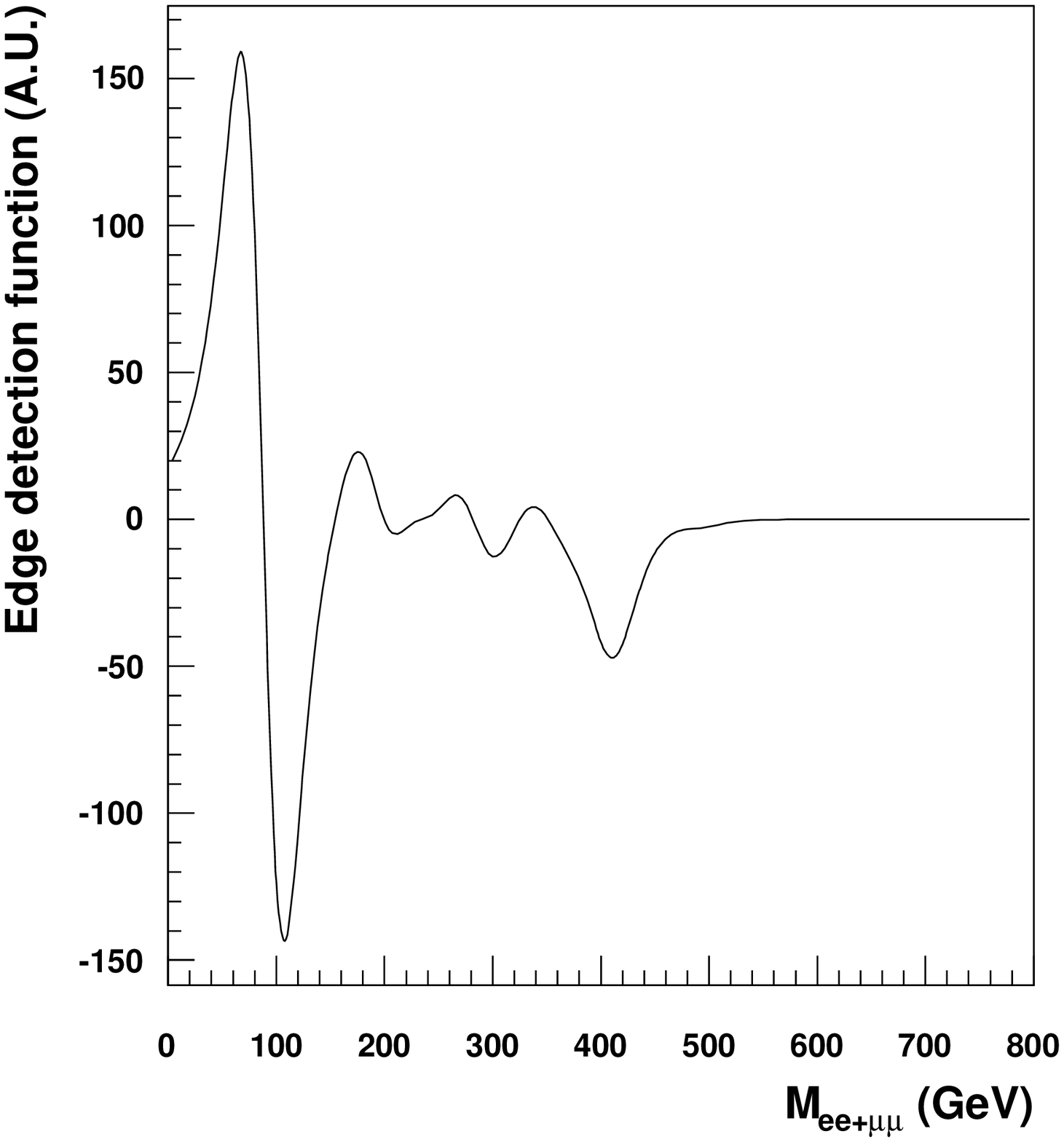} 
\label{fig:mll-100fb_edge}}
\subfigure[]{\includegraphics[width=0.4\textwidth]{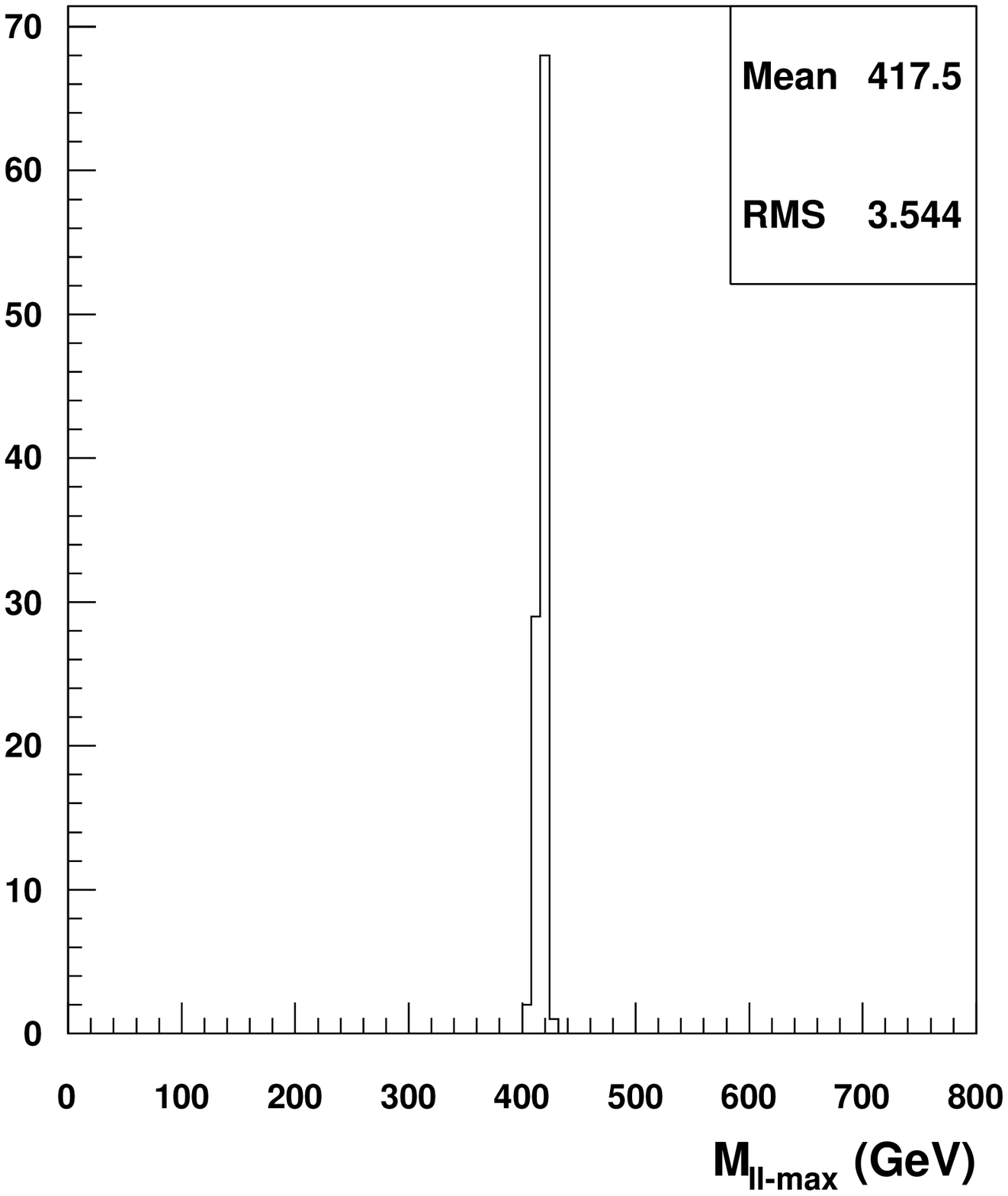} 
\label{fig:mllmax}}
\caption{(a) Histogram of events with two jets, two isolated leptons
  and missing energy after cuts (but including the $Z$ peak) as a
  function of $m_{\ell\ell}$ for the LNM--seq benchmark in a simulated
  experiment at the LHC with $\sqrt{s}$=14 TeV and assuming 100
  fb$^{-1}$ of integrated luminosity. The plot contains 726 events
  (502 applying the cut on the $Z$ peak). (b) Data filtering of the
  data in (a) through the function given in
  Eq.(\protect\ref{eq:filtering}), plotted as a function of the edge
  position guess and for a fixed value of the width parameter
  $\sigma=$ 30 GeV. The endpoint $m_{\ell\ell}^{max}$ of the
  distribution is represented by the rightmost minimum, and is
  reported on the data histogram in (a) with a vertical solid
  line. (c) Frequency histogram for the outcome of
  $m_{\ell\ell}^{max}$ for 100 random pseudo--experiments identical to
  the particular one plotted in (a) and (b).}
\end{figure}

Notice that Figs. \ref{fig:mll} and \ref{fig:mllj} are normalized to
the luminosity and represent the theoretical expectations of the
corresponding distributions. They are obtained using an integrated
luminosity of 546 $fb^{-1}$ (or 3940 events after selection), allowing
in particular to determine easily the position of the endpoints (that
agree with the values given in Table \ref{table:endpoints}). However,
for a lower value of the integrated luminosity the position of the
endpoints is blurred by statistical fluctuations, worsening their
determination. In the following we will assume for an optimistic and
yet realistic prediction of the latter quantity ${\cal L}$=100
fb$^{-1}$ at the end of the 14 TeV LHC run. The corresponding
prediction for the $m_{\ell\ell}$ distribution for a simulated
experiment is given in Fig. \ref{fig:mll-100fb} where the
$e\bar{e}$+$\mu\bar{\mu}$-$e\bar{\mu}-\mu\bar{e}$ subtraction between
same--flavor and different--flavor events has been applied to reduce
the SUSY background from chargino decays.  After selection cuts but
without subtracting the $Z$ peak this plot contains 726 events, which
become 502 when the $Z$ peak is subtracted.

The endpoint of this histogram provides the first edge
$m_{\ell\ell}^{max}$ needed for the kinematic reconstruction of the
masses. In order to find it we employ a method inspired by an
edge-detection algorithm frequently used in the field of image
processing and computer vision \cite{filtering}. Actually, the most
sensitive method for measuring the position of an edge would be to
obtain its expected distribution from a simulation and to perform a
likelihood fit to the data. However, this method depends on SUSY
parameters, hence lacking generality. On the other hand the
edge--detection algorithm, while it may not yield the best
sensitivity, has sufficient generality to be applied to a wider range
of problems.
   
The edge-detection algorithm is a method that allows to find the
endpoint of a sharply falling distribution by filtering the data
histogram through an appropriate function.  For concreteness, we take
as a filtering function $f(x,\mu,\sigma)=2\sinh\left( (x-\mu)/\sigma
\right)/\cosh^3 \left( (x-\mu)/\sigma \right)$ and try to minimize the
quantity:
\begin{equation}
  F(\mu,\sigma)=\sum_{i=1}^{N_{data}}f(x_i,\mu,\sigma),
\label{eq:filtering}
  \end{equation}

\noindent with respect to $\mu$ and with $\sigma$ fixed.  In
Eq.(\ref{eq:filtering}) $x_i$ represents the data count in the $i$--th
bin of the histogram. The width parameter $\sigma$ has the effect
of smoothing the distribution, hence making the algorithm immune to
noise. The choice of $\sigma$ is determined by looking at the width of
the distribution. If the value is too large, the edge determination is
imprecise, whereas if the value is too small, then it will be
sensitive to outliers.

In Fig. \ref{fig:mll-100fb_edge} we apply the method outlined above to
filter the $m_{\ell\ell}$ histogram of Fig.\ref{fig:mll-100fb}. We
assume $\sigma$=30 GeV and the endpoint position is represented by the
rightmost minimum, and is reported on the data histogram in
\ref{fig:mll-100fb} with a vertical solid line. In order to estimate
the statistical fluctuation of $m_{\ell\ell}^{max}$ we then repeat the
same procedure for 100 pseudo--experiments identical to the one
analyzed in Figs. \ref{fig:mll-100fb} and
\ref{fig:mll-100fb_edge}. The corresponding frequency histogram for
the outcome of $m_{\ell\ell}^{max}$ is given in Fig.\ref{fig:mllmax}.
In this way we find $m_{\ell\ell}^{max}$=417.5$\pm$ 3.5 (this value is
reported in parenthesis in Table \ref{table:endpoints}).

\begin{figure}
\subfigure[]{\includegraphics[width=0.5\textwidth,bb= 25 40 557 565]{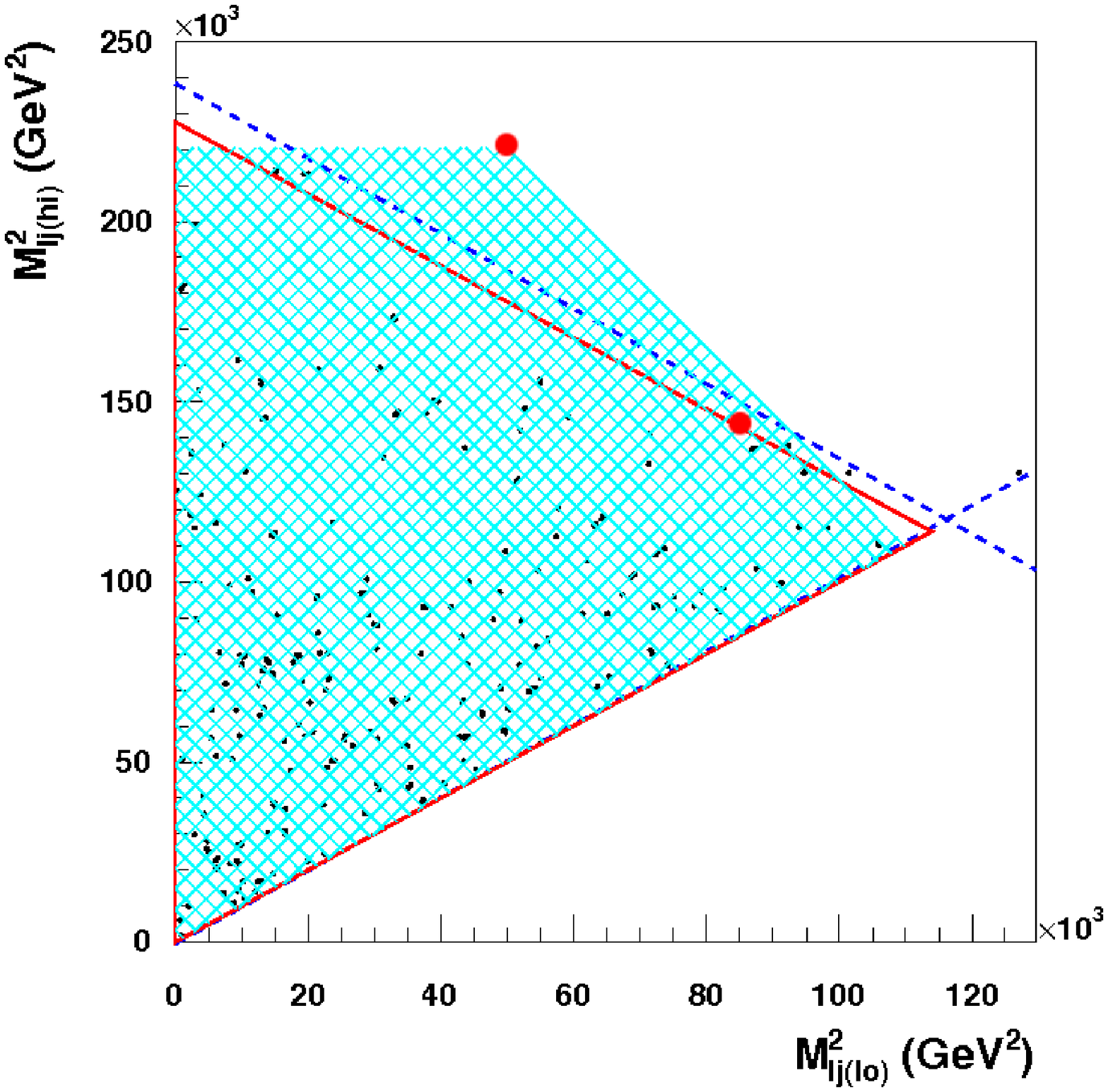}
\label{fig:mljhi_mljlo}}
\subfigure[]{\includegraphics[width=0.5\textwidth]{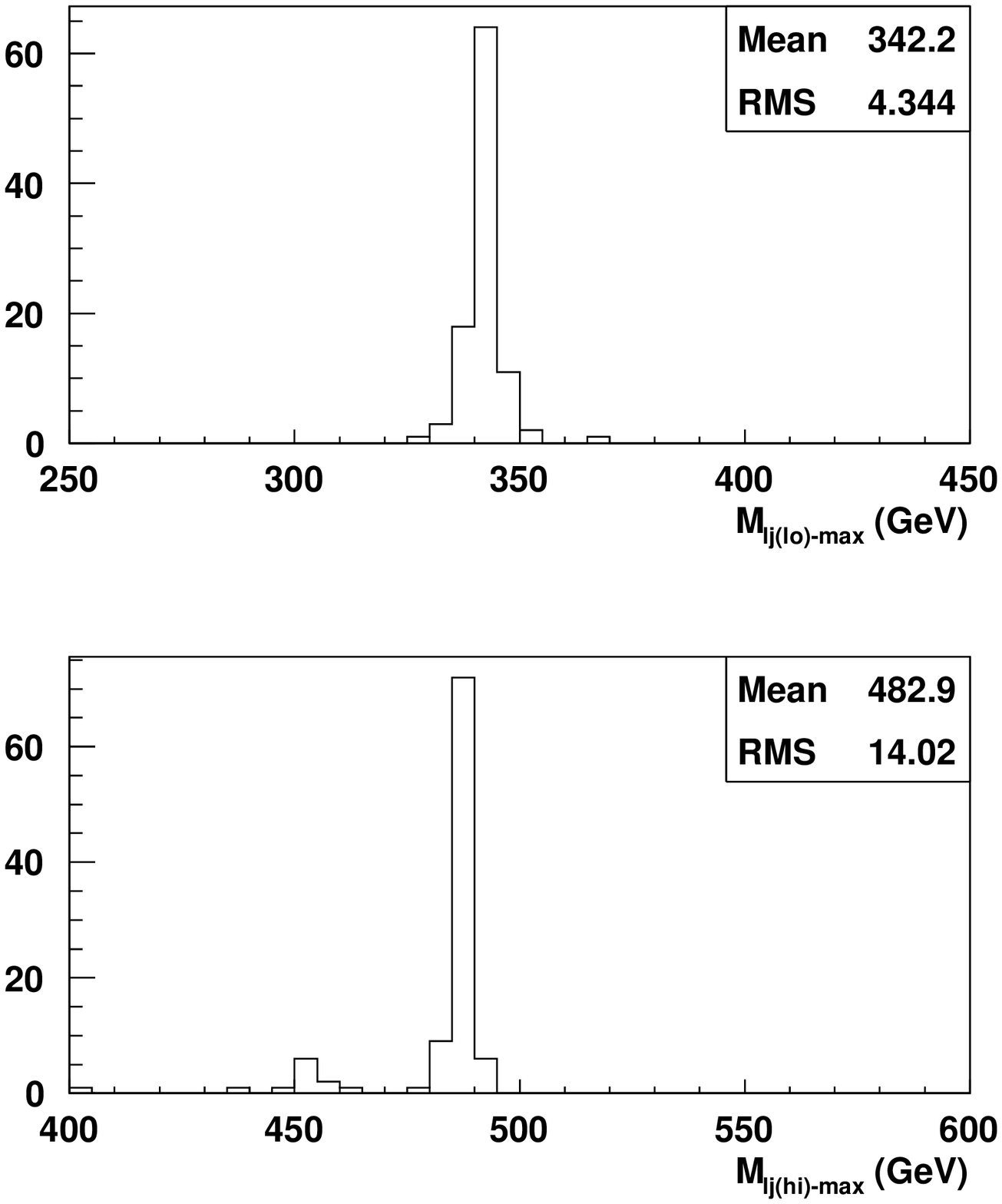}
\label{fig:mljhi_mljlo_edge}}
\caption{(a) Two--dimensional distribution in the plane
  $m_{(e+\mu)j(lo)}^2$--$m_{(e+\mu)j(hi)}^2$ of the events plotted in
  Fig.\protect\ref{fig:mll-100fb} that lie to the left of the value of
  $m_{\ell\ell}^{max}$ determined in
  Fig. \protect\ref{fig:mll-100fb_edge}, and when the cut on the $Z$
  peak is applied. The plot contains 497 events. The (red) solid line
  represents the expected boundary for LNM-seq, while the (blue)
  dashed lines are the fits of the boundaries, when the shape is
  assumed to be a triangular one (see text). The shaded area is the
  expected boundary for the duplicated model LNM-seq$^{\prime}$. The
  two filled circles show the positions of the point
  $(n^{\prime},p^{\prime})$ for LNM-seq and
  LNM-seq$^{\prime}$. Although for LNM-seq the exact position of
  $(n^{\prime},p^{\prime})$ cannot be measured, the fact that it lies
  on the boundary of the triangular shape is sufficient to break the
  degeneracy between the two duplicate mass patterns LNM--seq and
  LNM--seq$^{\prime}$ (see Table \protect\ref{table:endpoints} and
  text). (b) Frequency histogram for the output of the quantities
  $m_{j\ell (lo)}^{max}$ and $m_{j\ell (hi)}^{max}$ for 100 random
  pseudo--experiments identical to the particular one plotted in
  (a). In each pseudo--experiment $m_{j\ell (lo)}^{max}$ and $m_{j\ell
    (hi)}^{max}$ are obtained as the crossings of the fitted lines
  with the axes.}
\end{figure}

\begin{figure}
\subfigure[]{\includegraphics[width=0.5\textwidth]{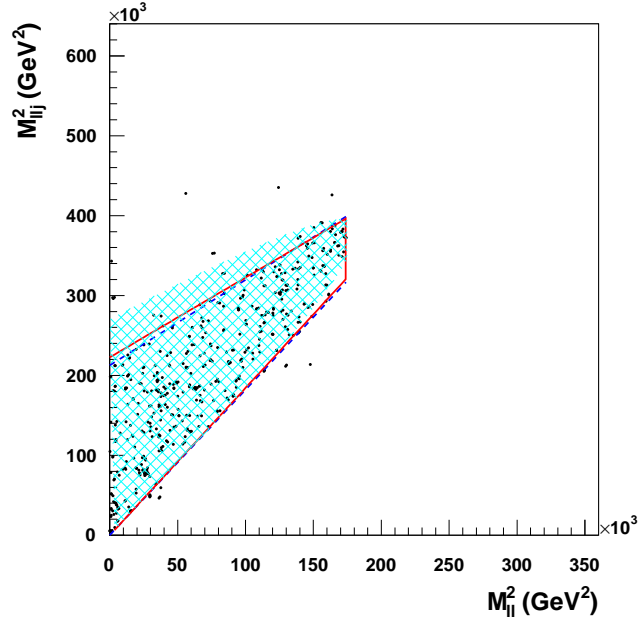} 
\label{fig:mllj_mll}}
\subfigure[]{\includegraphics[width=0.5\textwidth]{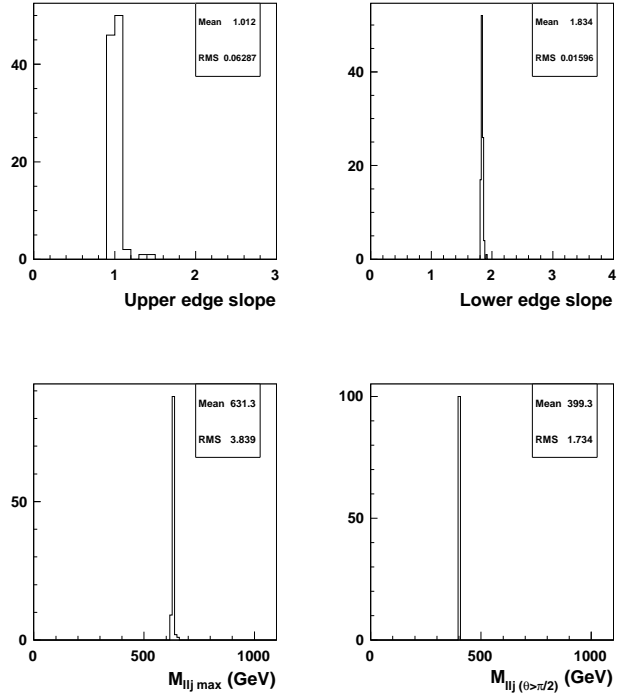} 
\label{fig:mllj_mll_edge}}
\caption{ (a) Two--dimensional distribution of the same events of
  Fig. \protect\ref{fig:mljhi_mljlo} in the plane
  $m_{ee+\mu\mu}^2$--$m_{(ee+\mu\mu)j}^2$. The (red) solid line
  represents the expected boundary.  The (blue) dashed lines are the
  fits of the upper and lower boundaries when they are assumed to be
  straight lines (see text). The value of $m_{\ell\ell}^{max}$ is the
  one determined in Fig. \protect\ref{fig:mll-100fb_edge}. The shaded
  area is the boundary for the duplicated model
  LNM-seq$^{\prime}$. (b) Frequency histogram for the outcome of the
  slopes of the upper and lower boundaries of the region in (a) and of
  the endpoints $m_{\ell\ell j}^{max}$ and $m_{\ell\ell j(\theta>\pi/2)}^{min}$, for 100 random pseudo--experiments identical to
  the particular one plotted in (a). The latter quantities are
  obtained as the crossing points of the relevant fitted boundary
  lines.\label{fig:mllj_mll_all}}
\end{figure}

In order to find the other endpoints needed to reconstruct the masses,
in principle the above procedure can be applied also to the histograms
obtained by plotting the same simulated events as a function of the
other invariant masses $m_{\ell\ell j}$, $m_{j\ell(hi)}$ and
$m_{j\ell(lo)}$. However, in the latter distributions the position of
the endpoints cannot be determined accurately because the number of
events is not large enough to saturate the endpoint of the histogram,
which systematically drops at a value considerably lower than the true
one for a lack of points in the tail. In this case an unambiguous
determination of the endpoint is strictly speaking impossible, and
only some educated guess can be made. In order to do this it can be
useful to resort to two--dimensional plots.  This is done in
Figs.\ref{fig:mljhi_mljlo} and \ref{fig:mllj_mll}, where the events of
the pseudo--experiment plotted in Fig.\ref{fig:mll-100fb}, and that
lie to the left of the determined value of $m_{\ell\ell}^{max}$, are
plotted in the planes \mjllo--\mjlhi and \mll--\mjll. In this way both
plots contain 497 events.

In the case of the \mjllo--\mjlhi two--dimensional plot of
Fig. \ref{fig:mljhi_mljlo} the shape of the region covered by the data
points nicely fits an isosceles triangle.  This very symmetric shape
is expected in two situations \cite{burns}: i) if the slepton is
produced off--shell (i.e. if $m_{\tilde{l}}>m_{\chi_i}$), since
in that case there is no longer distinction between the near and far
lepton and the two leptons have exactly the same kinematic properties;
ii) when the following relation among masses holds:

\begin{equation}
\frac{m_{\chi}^2}{m_{\tilde{l}}^2}<\frac{m_{\tilde{l}}^2}{m_{\chi_i}^2}<\frac{1}{2-m_{\chi}^2/m_{\tilde{l}}^2},
\label{eq:region_2}
\end{equation}

\noindent and at the same time there is a large hierarchy between the
slepton mass and the neutralino, $m_{\chi}/m_{\tilde{l}}\ll$1.  In the
latter case, that corresponds to our LNM-seq benchmark, the expected
boundary is actually delimited by four vertexes, but the fourth point
($n^{\prime},p^{\prime}$), that is supposed to be used to break the
degeneracy among duplicate models and whose coordinates are given in
Table \ref{table:endpoints}, lies on the straight line
$n^{\prime}+p^{\prime}$=(\mjll$^{max}$)$^2$ when $m_{\chi}\ll
m_{\tilde{l}}$ (it is represented by one of the two filled circles in
Fig.\ref{fig:mljhi_mljlo}) and cannot be observed
\footnote{The density of points of the two--dimensional distribution
  is expected to have a step--like drop for \mjllo$>p^{\prime}$
  \protect\cite{burns}, allowing in principle a determination of
  $p^{\prime}$. We have verified that in practice this measurement is
  not possible because of the large fluctuations in the determination
  of the density due to the low number of points.}.  Notice however
that, in spite of this, the degeneracy with the LNM-seq$^{\prime}$
model is easily broken, since the expected shape for the LNM-seq
benchmark, represented in Fig. \ref{fig:mljhi_mljlo} by the (red)
solid triangle, is very different from the corresponding one for
LNM-seq$^{\prime}$ shown as the shaded area in the same figure. As a
consequence of this the following relation among the masses:

\begin{equation}
0<\frac{m_{\tilde{l}}}{m_{\chi_i}}<\frac{m_{\chi}}{m_{\tilde{l}}},
\label{eq:region_3}
\end{equation}

\noindent which corresponds to the particular trapezoidal shape of the
LNM-seq$^{\prime}$ benchmark, can be safely discarded.

The distribution of points in Fig.\ref{fig:mljhi_mljlo} is clearly not
dense enough to saturate the vertexes of the triangle\footnote{If it
  were so the endpoints would be observable in the correspondent
  one--dimensional projections of the distribution, without the need
  to resort to two--dimensions in the first place!}, so strictly
speaking other kinematic regions different than the LNM-seq benchmark
cannot be ruled out (for a summary of shapes corresponding to
different kinematic situations see for instance Fig. 8 of
\cite{burns}). However the shape {\it is} very compatible to a
isosceles triangle as in the LNM-seq benchmark. Moreover in the
LNM-seq case the possibility that the triangular shape is due to an
off-shell sequential decay can be easily excluded on dynamical
grounds. In fact the branching ratio of the off--shell sequential
decay drops by at least two orders of magnitude compared to the
on--shell situation. In this case, in order to detect a few hundreds
events in the sequential channel as in Figs.\ref{fig:mljhi_mljlo},
\ref{fig:mllj_mll} the production cross section would need to be much
larger than in the LNM-seq, say in the range of a few tens pb. This in
turns would lead to a dramatic enhancement of branched decays that in
this case would be the dominant ones. In the LNM-seq scenario this
would lead to a huge number of events showing up in the $Z$ or Higgs
peaks when plotted as a function of $m_{\ell\ell}$. The non
observation of such an excess would easily allow to rule out that the
events plotted in Fig.\ref{fig:mljhi_mljlo} are due to off--shell
decays.

 In such a predicament we then propose to make the educated guess that
 the shape is a triangle and that the sequential decay is on shell.
 In this case, assuming that the boundaries of the region are straight
 lines, the edge-detection method that we used for the $m_{\ell\ell}$
 one--dimensional histogram can be modified to find the position of
 the edges in the two--dimensional $x$--$y$ plane by minimizing:
    \begin{equation}
  \pm\sum_{i=1}^{N_{data}}f(y_i-a*x_i-b,\mu,\sigma).
  \end{equation}
The sign '$\pm$' should be chosen depending on the observed slope of
the boundary. 

The result of the above procedure is shown in
Fig. \ref{fig:mljhi_mljlo}, where the fitted straight boundaries are
represented by the dashed (blue) lines. In this figure the two
corresponding endpoints $m_{j\ell (lo)}^{max}$, $m_{j\ell (hi)}^{max}$
are then obtained as the crossings of the boundaries with the two
axes. In Fig.\ref{fig:mljhi_mljlo_edge} the same procedure is repeated
for 100 pseudo--experiments identical to the one shown in
Fig.\ref{fig:mljhi_mljlo}, and the frequency histogram for the output
values of $m_{j\ell (lo)}^{max}$ and $m_{j\ell (hi)}^{max}$ are
given. The corresponding determination for $m_{j\ell (lo)}^{max}$ and
$m_{j\ell (hi)}^{max}$ is reported in parenthesis in Table
\ref{table:endpoints}.

The determination of the last two endpoints $m_{j\ell\ell}^{max}$ and
$m_{j\ell\ell(\theta<\pi/2)}^{min}$ is finally discussed in
Fig. \ref{fig:mllj_mll_all}. As in the previous figure the (red) solid
line represents the expected boundary for the LNM-seq benchmark, while
the shaded area is the corresponding one for the duplicate model
LNM-seq$^{\prime}$.  Also in this case the upper and lower boundaries
of the region covered by the simulated data are compatible with
straight lines. This is broadly consistent with the guess that the
kinematic region of Fig.\ref{fig:mllj_mll} is due to an on-shell decay
\cite{burns}. When the boundaries are intersected with the value of
$m_{\ell\ell}^{max}$ determined in Fig.\ref{fig:mll-100fb_edge}, both
$m_{j\ell\ell}^{max}$ and $m_{j\ell\ell(\theta<\pi/2)}^{min}$ can be
obtained. Finally, in Fig.\ref{fig:mllj_mll_edge} repeating the same
procedure for 100 pseudo--experiments identical to that of
Fig.\ref{fig:mllj_mll} the frequency histograms for the output values
of $m_{j\ell\ell}^{max}$, $m_{j\ell\ell(\theta<\pi/2)}^{min}$, and of the
slopes of the upper and lower boundaries is obtained. The ensuing
determinations of $m_{j\ell\ell}^{max}$ and
$m_{j\ell\ell(\theta<\pi/2)}^{min}$ are given in parenthesis in Table
\ref{table:endpoints}.

\begin{figure}
\includegraphics[width=0.7\textwidth,bb= 25 37 564 620]{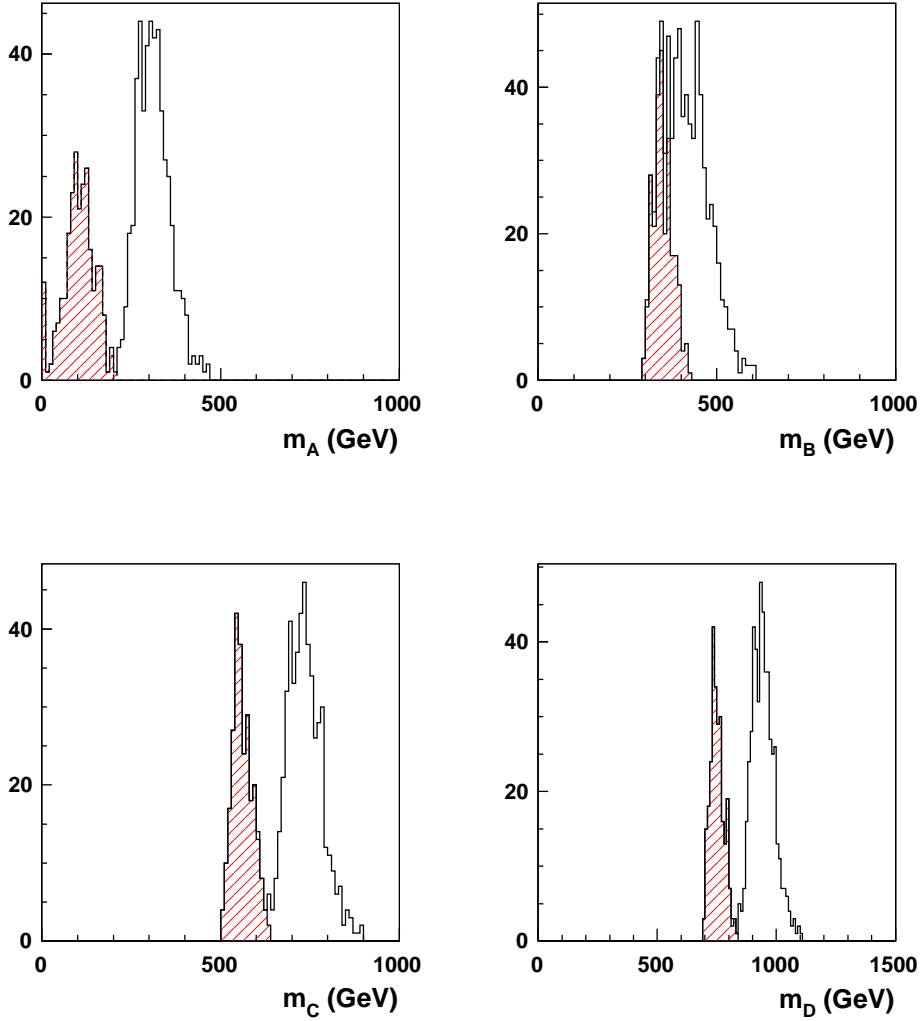}
\caption{Mass determination for the particles of the sequential decay
  obtained from the values of the endpoints measured using the
  simulated sample of Figs.  \protect\ref{fig:mljhi_mljlo} and
  \protect\ref{fig:mllj_mll}. Upper left: $m_{\chi}$; upper right:
  $m_{\tilde{l}}$; lower left: $m_{\chi_i}$; lower right:\msq. In all
  plots the histogram bins are filled with a shaded box in
  correspondence to mass patterns verifying
  Eq. (\protect\ref{eq:region_2}). Only the mass determinations
  corresponding to the shaded peaks are compatible to the data, namely
  to the shape of the data region of Fig.\protect\ref{fig:mljhi_mljlo}
  in the \mjlhi--\mjllo plane.
   \label{fig:massfit}}
\end{figure}

Once the 5 endpoints $m_{\ell\ell}^{max}$, $m_{j\ell(hi)}^{max}$,
$m_{j\ell(low)}^{max}$, $m_{j\ell\ell}^{max}$,
$m_{j\ell\ell(\theta<\pi/2)}^{min}$ are determined (notice that, as
mentioned before, since the relation of Eq. (\ref{eq:correlation})
holds, only four of them are independent) they can be used to
determine the masses. The mass inversion is obtained in a
straightforward way by simulating a large number of random values of
the four masses $m_{\chi}$, $m_{\tilde{l}}$, $m_{\chi_i}$,
$m_{\tilde{q}}$ and plotting the histogram of the mass combinations
whose theoretical values of the endpoints fall within the measured
ranges. The result of such an inversion is shown in
Fig. \ref{fig:massfit}. As expected, this procedure leads to two
different solutions, corresponding to the LNM-seq benchmark and to the
duplicate one LNM-seq$^{\prime}$. Notice however that from the
discussion of Fig.\ref{fig:mljhi_mljlo} this degeneracy can be easily
broken.  In fact the distribution of the simulated data points in the
\mjllo--\mjlhi plane is strongly inconsistent with the
LNM-seq$^{\prime}$ solution, allowing to conclude that only mass
patterns verifying Eq.(\ref{eq:region_2}) are compatible with the
simulated data. In Fig.\ref{fig:massfit} such mass patterns are marked
by filling the bins with a shaded box. From this figure one can see
that the correct solution can be clearly discriminated from the
duplicate one for all the masses involved the decay.  In this way from
Fig. \ref{fig:massfit} we get the following determination of the
masses:

\begin{eqnarray}
m_{\chi}&=& (103 \pm 43) \,\,{\rm GeV}   \label{eq:final_mchi}\\
m_{\tilde{l}}&=& (349 \pm 27) \,\,{\rm GeV}   \label{eq:final_m_sl} \\
m_{\chi_i}&=&  (561 \pm 28) \,\,{\rm GeV}\label{eq:final_mchi_prime}  \\
m_{\tilde{q}}&=& (751 \pm 28) \,\,{\rm GeV} \label{eq:final_m_sq} .
\end{eqnarray}

The reconstructed value for the neutralino mass deviates remarkably
from its input value. This is explained by the fact that the
center--of--mass energy available in the sequential decays is set by
the squark and is much larger than the neutralino mass. So neutralinos
are produced in the relativistic regime, in which their kinematics is
almost insensitive to the actual value of $m_{\chi}$. Moreover, the
dependence of the kinematic endpoints on the neutralino mass is
quadratic, so determining $m_{\chi}$ implies taking a square root,
which worsens the accuracy compared to that with which the endpoints
are measured. However, the reconstructed value for $m_{\chi}$, though
deviated from the actual value, would entail the important indication
that some neutral stable particle is being produced in the process,
whose mass can be compatible to a light range. Notice that the similar
yield to leptons of the three families observed in Figs.~\ref{fig:mll}
and \ref{fig:mllj} allows to conclude, as discussed in
Section~\ref{sec:properties}, that both the neutralino and the
$\chi_i$ particle are of gaugino type. As explained in
Section~\ref{sec:model} this implies $m_{\chi}\simeq M_1$ and
$m_{\chi_i}\simeq M_2$, so the reconstructed mass spectrum would be
incompatible with neutralinos in SUGRA scenarios, since it would point
toward a ratio $M_1/M_2\lsim 0.5$, in disagreement with what expected
in models where gaugino masses are unified at the GUT scale.

\section{Conclusions}
\label{sec:conclusions}
Signals of relic particles in direct DM searches raise the interest
for masses of these relic particles in the range 7-8 GeV $\lsim
m_{\chi} \lsim$ 50 GeV.  This is actually the mass range compatible
with the annual-modulation effect measured by the DAMA collaboration
\cite{dama2010}, when this effect is interpreted in terms of DM
particles with an elastic coherent interaction with nuclei.  If the
positive results of other experiments of DM direct detection are taken
into account (CDMS \cite{cdms}, CoGeNT \cite{cogent}, CRESST
\cite{cresst}) the mass range restricts to 7-8 GeV $\lsim m_{\chi}
\lsim$ 15-20 GeV \cite{noi_belli_et_al_2011}.  These experimental
results are fitted quite well by the Light Neutralino Model
\cite{lowneu,noi_discussing} that is an effective Minimal
Supersymmetric extension of the Standard Model at the electroweak
scale without requirement of a gaugino-mass unification at a grand
unification scale.  At variance with Supergravity--inspired (SUGRA)
models, in the LNM the neutralino mass can be as small as about 9 GeV,
as discussed in the Introduction.

In view of the interest of these light neutralinos in the
phenomenology related to DM direct detection, in the present paper we
have addressed the problem of a search at the LHC for a neutralino of
very light mass.  A preliminary analysis in this direction was
performed in Ref.~\cite{lhc1}. There, specific scenarios and
benchmarks within the LNM and dictated by relevant cosmological
properties were considered, and the relevant expected event rates
determined; however, no specific analyzes of the signal/background
ratios and of kinematical distributions were performed.  

In the present paper we have extended the investigation of
Ref.~\cite{lhc1} by making use of numerical simulations to estimate in
a realistic way the detectability of light neutralinos at the LHC over
the SM background and to show what information about the masses of
SUSY particles can be extracted from the data.  Within the sequential
and branched decay chains that constitute the typical processes by
which neutralinos can be searched for at the LHC, we have singled out
the most dominant contributions in the context of the LNM. For this
aspect, the analytical expressions for the light neutralino
spectroscopy reported in Sect. II have been exploited.

We have then selected a benchmark (dubbed LNM-seq) with the specific
feature of belonging to a value (around 10 GeV) of the neutralino mass
that is representative of the neutralino mass values in the low-side
of the $m_{\chi}$ ranges mentioned above.  In terms of this benchmark
we have investigated which are the expectations for having some
signals at the LHC at different stages of the LHC operation.

We have found that with the integrated luminosity ${\cal L}\simeq$5
fb$^{-1}$ that would be collected at the end of the 2011 run at a
center-of-mass energy of 7 TeV, the LNM-seq benchmark is expected to
provide a slight excess over the background, namely at the level of a
$\simeq$ 3.2 $\sigma$ significance, assuming that our estimation on
the background has a 5\% relative uncertainty.  As discussed in
Sect.~\ref{seq:early}, this would not be sufficient to draw any
conclusions on the mass and properties of the neutralino.

We have then analyzed the prospects in terms of the integrated
luminosity of ${\cal L}\simeq$100 fb$^{-1}$ that might be reached by
the LHC at the end of its 14 TeV run.  To this purpose a detailed
analysis has been performed by employing one-dimensional and
two-dimensional distributions with the scope of establishing how the
inputs of the LNM-seq benchmark can be reconstructed by the
determinations of the relevant end-points in the various mass
distributions. The problem of the disentanglement of the true solution
from the duplicate solution in the inversion procedure from the
end-point values to the model parameters has been addressed.  

The main result concerns the reconstruction of the neutralino mass
that finally turns out to be determined as $(m_{\chi})_ {rec} = 103
\pm 43$ GeV. This value deviates remarkably from the input value for
$m_{\chi}$, but this result is not surprising in view of the
difficulty in reconstructing the mass of light stable particles in
relativistic events. However, the reconstructed value for $m_{\chi}$,
though deviated from the actual value, would entail the important
indication that some neutral stable particle is being produced in the
process, whose mass can be compatible to a light range.  Moreover, the
reconstructed masses would suggest $M_1$/$M_2\lsim$0.5, in
disagreement with what expected in SUGRA models where gaugino masses
are unified at the GUT scale.

\acknowledgments

A.B. and N.F. acknowledge Research Grants funded jointly by Ministero
dell'Istruzione, dell'Universit\`a e della Ricerca (MIUR), by
Universit\`a di Torino and by Istituto Nazionale di Fisica Nucleare
within the {\sl Astroparticle Physics Project} (MIUR contract number:
PRIN 2008NR3EBK; INFN grant code: FA51). S.S. acknowledges support by
NRF with CQUEST grant 2005-0049049 and by the Sogang Research Grant
2010. N.F. acknowledges support of the spanish MICINN Consolider
Ingenio 2010 Programme under grant MULTIDARK CSD2009- 00064.
S.C. acknowledges support from the Korean National Research Foundation
NRF-2010-0015467.
\medskip

\medskip

\appendix

\section{Analytic expressions of kinematic endpoints}
\label{sec:analytic_endpoints}

We give here for completeness the analytic expressions used to
calculate the kinematic endpoints in Table\ref{table:endpoints}. These
formulae are taken from Ref.\cite{burns}.

\begin{eqnarray}
  \left(m_{\ell\ell}^{max}\right)^2 &=& m_{\tilde{q}}^2\, R_{\chi_i\tilde{q}}\, 
(1-R_{\tilde{l}\chi_i})\, (1-R_{\chi\tilde{l}}); 
\label{eq:ll_def} \\ [4mm]
\left(m_{j\ell\ell}^{max}\right)^2 &=& 
\left\{ 
\begin{array}{ll}
m_{\tilde{q}}^2 (1-R_{\chi_i\tilde{q}})(1-R_{\chi\chi_i}), & ~{\rm for}\ R_{\chi_i\tilde{q}}<R_{\chi\chi_i},       \\[4mm] 
m_{\tilde{q}}^2 (1-R_{\tilde{l}\chi_i})(1-R_{\chi\tilde{l}}R_{\chi_i\tilde{q}}), 
& ~{\rm for}\ R_{\tilde{l}\chi_i}<R_{\chi\tilde{l}}R_{\chi_i\tilde{q}}, \\[4mm] 
m_{\tilde{q}}^2 (1-R_{\chi\tilde{l}})(1-R_{\tilde{l}\tilde{q}}),       & ~{\rm for}\ R_{\chi\tilde{l}}<R_{\tilde{l}\tilde{q}},       \\[4mm] 
m_{\tilde{q}}^2\left(1-\sqrt{R_{\chi\tilde{q}}}\,\right)^2,   & ~{\rm otherwise}.
\end{array} \right . \label{eq:jll_def} \\
\left(m_{j\ell(lo)}^{max}\right)^2&=&\left\{ 
\begin{array}{ll}
\left(m_{j\ell_n}^{max}\right)^2,   & ~{\rm for}\ (2-R_{\chi\tilde{l}})^{-1} < R_{\tilde{l}\chi_i} < 1, \\[4mm] 
\left(m_{j\ell(eq)}^{max}\right)^2, & ~{\rm for}\ R_{\chi\tilde{l}}< R_{\tilde{l}\chi_i}<(2-R_{\chi\tilde{l}})^{-1}, \\[4mm] 
\left(m_{j\ell(eq)}^{max}\right)^2, & ~{\rm for}\ 0< R_{\tilde{l}\chi_i}<R_{\chi\tilde{l}},              
\end{array}%
\right . \label{eq:jl_hi_def} \\
\left( m_{j\ell(hi)}^{max}\right)^2 &=&\left\{ 
\begin{array}{ll}
\left(m_{j\ell_f}^{max}\right)^2, & ~{\rm for}\ (2-R_{\chi\tilde{l}})^{-1} < R_{\tilde{l}\chi_i} < 1, \\[4mm] 
\left(m_{j\ell_f}^{max}\right)^2, & ~{\rm for}\ R_{\chi\tilde{l}}< R_{\tilde{l}\chi_i}<(2-R_{\chi\tilde{l}})^{-1}, \\[4mm] 
\left(m_{j\ell_n}^{max}\right)^2, & ~{\rm for}\ 0< R_{\tilde{l}\chi_i}<R_{\chi\tilde{l}},   \label{eq:jl_lo_def}            
\end{array}%
\right .
\end{eqnarray}

\begin{eqnarray}
\left( m_{j\ell\ell(\theta>\frac{\pi}{2})}^{min}\right)^2 &=& 
\frac{1}{4}m_{\tilde{q}}^2 \Biggl\{ (1-R_{\chi\tilde{l}})(1-R_{\tilde{l}\chi_i})(1+R_{\chi_i\tilde{q}})  \label{eq:jlltheta} 
\\ \nonumber
&+& 2\, (1-R_{\chi\chi_i})(1-R_{\chi_i\tilde{q}})
-(1-R_{\chi_i\tilde{q}})\sqrt{(1+R_{\chi\tilde{l}})^2 (1+R_{\tilde{l}\chi_i})^2-16 R_{\chi\chi_i}}\Biggr\} ,
\end{eqnarray}

\noindent with:

\begin{eqnarray}
\left(m_{j\ell_n}^{max}\right)^2   &=& m_{\tilde{q}}^2\, (1-R_{\chi_i\tilde{q}})\, (1-R_{\tilde{l}\chi_i})\, , \label{mjlnmax}\\
\left(m_{j\ell_f}^{max}\right)^2   &=& m_{\tilde{q}}^2\, (1-R_{\chi_i\tilde{q}})\, (1-R_{\chi\tilde{l}})\, , \label{mjlfmax}\\
\left(m_{j\ell(eq)}^{max}\right)^2 &=& m_{\tilde{q}}^2\, (1-R_{\chi_i\tilde{q}})\, (1-R_{\chi\tilde{l}})\, (2-R_{\chi\tilde{l}})^{-1} \, , \label{mjleqmax}
\end{eqnarray}

\noindent and $R_{lm}\equiv m_l^2/m_m^2$ with $l=\chi,\tilde{l},\chi_i,\tilde{q}$. 

Moreover, the quantities $(n^{\prime},p^{\prime})$ are given by:

\begin{eqnarray}
n^{\prime}&=&min(n,p) \nonumber \\
p^{\prime}&=&max(n,p), \label{eq:n_prime_p_prime}
\end{eqnarray}

\noindent where:

\begin{eqnarray}
n&=& \left(m_{j\ell_n}^{max}\right)^2= m^2_{\tilde{q}} \left ( 1-R_{\chi_i \tilde{q}} \right) \left ( 1-R_{\tilde{l}\chi_i} \right)   \nonumber \\
p&=& \left(m_{j\ell_f}^{max}\right)^2= m^2_{\tilde{q}} \left ( 1-R_{\chi_i \tilde{q}} \right) \left ( 1-R_{\chi\tilde{l}} \right). \label{eq:np}
\end{eqnarray}



\end{document}